\documentclass{aastex62}

\graphicspath{{./}{figures/}}

\accepted{February 24, 2020}

\shorttitle{\nth{4} \Fermi-GBM Gamma-Ray Burst Catalog}
\shortauthors{von Kienlin et al.}

\usepackage{hyperref}
\usepackage{csquotes}
\usepackage{graphicx}
\usepackage{epstopdf}
\usepackage{lineno}
\usepackage{nicefrac} 
\usepackage{amsmath,amstext,mathrsfs}
\usepackage[super]{nth}
\usepackage{comment}
\usepackage{xspace}
\usepackage{soul}
\usepackage{xfrac}
\newcommand{\Fermi}{\textit{Fermi\xspace}}
\newcommand{\Swift}{\textit{Swift\xspace}}
\newcommand{\INTEGRAL}{\textit{INTEGRAL\xspace}}
\newcommand{\Suzaku}{\textit{Suzaku\xspace}}
\newcommand{\AGILE}{\textit{AGILE\xspace}}
\newcommand{\MAXI}{\textit{MAXI\xspace}}
\newcommand{\RHESSI}{\textit{RHESSI\xspace}}
\newcommand{\NuSTAR}{\textit{NuSTAR\xspace}}
\newcommand{\Messenger}{\textit{Messenger\xspace}}
\newcommand{\tn}{$T_{\rm \,90}$}
\newcommand{\tf}{$T_{\rm \,50}$}

\begin{document}

\title{The Fourth \Fermi-GBM Gamma-Ray Burst Catalog:  A Decade of Data}

\correspondingauthor{A.~von Kienlin}
\email{azk@mpe.mpg.de}
\author[0000-0002-0221-5916]{A.~von Kienlin}
\affil{Max-Planck-Institut f\"{u}r extraterrestrische Physik, Giessenbachstrasse 1, D-85748 Garching, Germany}
\author{C.~A.~Meegan}
\affil{Center for Space Plasma and Aeronomic Research, University of Alabama in Huntsville, 320 Sparkman Drive, Huntsville, AL 35899, USA}
\author[0000-0002-2481-5947]{W.~S.~Paciesas}
\affiliation{Science and Technology Institute, Universities Space Research Association, 320 Sparkman Drive, Huntsville, AL 35805, USA}
\author[0000-0001-7916-2923]{P.~N.~Bhat}
\affil{Center for Space Plasma and Aeronomic Research, University of Alabama in Huntsville, 320 Sparkman Drive, Huntsville, AL 35899, USA}
\affil{Department of Space Science, University of Alabama in Huntsville, 320 Sparkman Drive, Huntsville, AL 35899, USA}
\author[0000-0001-9935-8106]{E.~Bissaldi}
\affil{Dipartimento Interateneo di Fisica, Politecnico di Bari, Via E. Orabona 4, I-70125 Bari, Italy}
\affil{INFN - Sezione di Bari, Via E. Orabona 4, I-70125 Bari, Italy}
\author{M.~S.~Briggs}
\affil{Center for Space Plasma and Aeronomic Research, University of Alabama in Huntsville, 320 Sparkman Drive, Huntsville, AL 35899, USA}
\author{E.~Burns}
\affiliation{NASA Postdoctoral Program Fellow, Goddard Space Flight Center, Greenbelt, MD 20771, USA}
\author{W.~H.~Cleveland}
\affil{Science and Technology Institute, Universities Space Research Association, 320 Sparkman Drive, Huntsville, AL 35805, USA}
\author{M.~H.~Gibby}
\affiliation{Jacobs Space Exploration Group, Huntsville, AL 35806, USA}
\author{M.~M.~Giles} 
\affiliation{Jacobs Space Exploration Group, Huntsville, AL 35806, USA}
\author[0000-0002-0587-7042]{A.~Goldstein}
\affiliation{Science and Technology Institute, Universities Space Research Association, 320 Sparkman Drive, Huntsville, AL 35805, USA}
\author{R.~Hamburg}
\affil{Department of Space Science, University of Alabama in Huntsville, 320 Sparkman Drive, Huntsville, AL 35899, USA}
\affil{Center for Space Plasma and Aeronomic Research, University of Alabama in Huntsville, 320 Sparkman Drive, Huntsville, AL 35899, USA}
\author[0000-0002-0468-6025]{C.~M.~Hui}
\affiliation{NASA Marshall Space Flight Center, Huntsville, AL 35812, USA}
\author{D.~Kocevski}
\affiliation{Astrophysics Office, ST12, NASA/Marshall Space Flight Center, Huntsville, AL 35812, USA}
\author{B.~Mailyan}
\affil{Center for Space Plasma and Aeronomic Research, University of Alabama in Huntsville, 320 Sparkman Drive, Huntsville, AL 35899, USA}
\author[0000-0002-0380-0041]{C.~Malacaria}
\affiliation{NASA Postdoctoral Program Fellow,  Marshall Space Flight Center, NSSTC, 320 Sparkman Drive, Huntsville, AL 35805, USA}
\affiliation{Universities Space Research Association, NSSTC, 320 Sparkman Drive, Huntsville, AL 35805, USA}
\author[0000-0002-6269-0452]{S.~Poolakkil}
\affil{Department of Space Science, University of Alabama in Huntsville, 320 Sparkman Drive, Huntsville, AL 35899, USA}
\affil{Center for Space Plasma and Aeronomic Research, University of Alabama in Huntsville, 320 Sparkman Drive, Huntsville, AL 35899, USA}
\author[0000-0003-1626-7335]{R.~D.~Preece}
\affil{Department of Space Science, University of Alabama in Huntsville, 320 Sparkman Drive, Huntsville, AL 35899, USA}
\author[0000-0002-7150-9061]{O.~J.~Roberts}
\affil{Science and Technology Institute, Universities Space Research Association, 320 Sparkman Drive, Huntsville, AL 35805, USA}
\author[0000-0002-2149-9846]{P.~Veres}
\affiliation{Center for Space Plasma and Aeronomic Research, University of Alabama in Huntsville, 320 Sparkman Drive, Huntsville, AL 35899, USA}
\author[0000-0002-8585-0084]{C.~A.~Wilson-Hodge}
\affiliation{NASA Marshall Space Flight Center, Huntsville, AL 35812, USA}



\begin{abstract}
We present the fourth in a series of catalogs of gamma-ray bursts (GRBs) observed with \Fermi's Gamma-Ray Burst Monitor (\Fermi-GBM). It extends the six year catalog by four more years, now covering the 10 year time period from trigger enabling on 2008 July 12 to 2018 July 11. During this time period GBM triggered almost twice a day on transient events of which we identified 2356 as cosmic GRBs. 
Additional trigger events were due to  solar flare events, magnetar burst activities, and terrestrial gamma-ray flashes.
The intention of the GBM GRB catalog series is to provide updated information to the community on the most important observables of the GBM-detected GRBs.  For each GRB the location and main characteristics of the prompt emission, the duration, peak flux, and fluence are derived. The latter two quantities are calculated for the 50--300~keV energy band, where the maximum energy release of GRBs in the instrument reference system is observed and also for a broader energy band from 10--1000~keV, exploiting the full  energy range of GBM's low-energy detectors. Furthermore, information is given on the settings of the triggering criteria and exceptional operational conditions during years seven to ten in the mission. This fourth catalog is an official product of the \Fermi-GBM science team, and the data files containing the complete results are available from the High-Energy Astrophysics Science Archive Research Center (HEASARC).

\end{abstract}


\section{Introduction} \label{sec:intro}

With the completion of the first decade of operation, \Fermi-GBM has been in orbit longer than its predecessor experiment, the Burst and Transient Source Experiment (BATSE)\footnote{\url{https://gammaray.nsstc.nasa.gov/batse/}}  on board the {\it Compton Gamma Ray Observatory} (CGRO, $ \sim 9$ year of operation). Despite its lower sensitivity and smaller detectors, the GBM instrument is capable of detecting  almost the same number of GRBs ($\sim 240$ GBM GRBs  compared to $\sim 300$ BATSE GRBs per year) mostly thanks to its advanced triggering system \citep{2012ApJS..199...18P}. Thus it successfully continues to detect and coarsely locate GRBs over a wide field of view (FOV), and to provide broad spectral information in the hard X-ray and soft gamma-ray energy range (8~keV--40~MeV) where bursts emit most of their energy.

The \Fermi-GBM science team releases catalogs on a regular basis that list the main characteristics of triggered bursts,
compiling the data of several completed mission years. These have included the first two \citep[\nth{1} GBM GRB catalog,][]{2012ApJS..199...18P}, four \citep[\nth{2},][]{2014ApJS..211...13V}, and six \citep[\nth{3},][]{2016ApJS..223...28N} mission years, which are now continued by the current 10 year catalog. The first two catalogs were accompanied by spectral catalogs, for the first two \citep{2012ApJS..199...19G} and four  \citep{2014ApJS..211...12G} mission years, which provide more detailed information on the spectral characteristics of nearly all GRBs, including the time-integrated fluence and peak flux spectra.
These results are updated by the current 10 year spectral catalog (S.~Poolakkil et al. 2020, in preparation).
A time-resolved spectral analysis of the brightest 81 GRBs observed during the first four mission years is provided in the first time-resolved spectral catalog \citep{2016A&A...588A.135Y}. It will be continued in a forthcoming catalog (E. Bissaldi et al., in preparation) presenting the time-resolved spectral analysis for the brightest GRBs of the first 10 years. 

The GRB detection capabilities of GBM are augmented by \Fermi's primary instrument, the Large Area Telescope (LAT), which overlaps and extends the GBM energy range (30~MeV--300~GeV), allowing observations over more than seven decades in energy. 
The second LAT GRB catalog \citep{2019ApJ...878...52A}, which covers the first 10 year of operations, from 2008 to 2018 August 4, lists 176 GRBs jointly detected by LAT and GBM, emphasizing the great scientific merit of LAT 
in uncovering previously unknown characteristics of GRBs at high gamma-ray energies. Examples include the delayed onset and extended duration of the emission above 100~MeV and the observation of additional spectral components. 
We note that LAT detected an additional 10 GRBs that were
independently detected by instruments other than GBM.

\begin{table}[b]
  \centering
  \caption{Trigger statistics of the first 10 mission years, subdivided into  2 Year sections. \label{tab:trigstat4thcat}}
\begin{tabular}{clcccccccccc}
  \hline \hline
 Cat \# & Year\tablenotemark{a}  & GRBs & SGRs & TGFs & SFs & Galactic & CPs & Other & Sum & ARRs\tablenotemark{d} & LAT GRBs \\ \hline
1 &  1 to 2 & 494\tablenotemark{b}  & 150 & 79 & 29 & 4 & 55& 52& 862\tablenotemark{c} & 40 & 38\\
2 &  3 to 4 & 466 & 17 &183 & 363 & 0 & 132 & 59& 1220 & 47 & 29\\
3 &  5 to 6 & 451\tablenotemark{b}  & 9 &207 &399& 2 &90 & 65 & 1223& 33 & 42 \\
4 &  7 to 8 & 464\tablenotemark{b} & 65 &215 &318& 173 & 422& 82 & 1739& 47 &  36\\
4 &  9 to 10 & 485 & 17 &196 &67& 228 & 324 & 73 & 1390 & 53 & 41\\
4 & 1 to  10 & 2360\tablenotemark{b} & 258 & 880 & 1176& 407 & 1023& 331 & 6434 & 220\tablenotemark{e} & 186 \\
  \hline
\end{tabular}
\tablenotetext{a}{The triggers of a mission year are always counted from July 12 to  July 11 of the following year, starting with trigger enabling on 2008, July 12.}
\tablenotetext{b}{GRB 091024A, GRB 130925A, GRB 150201A and GRB 160625B each of which triggered GBM twice, are counted twice. Hence, the total number of GRB's is one less in mission years 1 and 2 and  5 and 6, two less in missions years 7 and 8 and four less for the ten year sum.}
\tablenotetext{c}{The total numbers of triggers is two less compared to \cite{2012ApJS..199...18P}, since the two commanded triggers (bn100709294 \& bn100711145) were not counted.}
%
\tablenotetext{d} {derived from the \Fermi~timeline posting page at FSSC: \url{https://fermi.gsfc.nasa.gov/ssc/observations/timeline/posting/arr/} }
\tablenotetext{e} {Due to misclassification of events as GRBs by the flight software (FSW), 48 of the ARRs occurred for other event types. Of these, 34 occurred due to charged particle events, 5 occurred due to SGR events, 8 occurred due to solar flare events, and 1 was due to a TGF event. In addition, there were a few positive ARRs from GBM triggers followed by no spacecraft slews, which were disabled at the spacecraft level at that time. In a few cases, the spacecraft slew started well after the GBM trigger due to Earth's limb constraint.}
\end{table}

In addition to the standard \Fermi-GBM GRB catalog products, which are the GRB location, duration, peak flux, and fluence, the first three catalogs provided supplementary information. \cite{2012ApJS..199...18P} investigated the apparent improvement in trigger sensitivity relative to BATSE, which was discussed in more detail in the second catalog \citep{2014ApJS..211...13V}, including a comparison of the numbers of BATSE- and non-BATSE-like GBM GRB triggers.  The six year catalog \citep{2016ApJS..223...28N} provided an accurate estimate of the daily burst rate and employed statistical methods to assess clustering in the GRB duration-hardness distribution.  It was found that the GRBs are better fit by a two-component model with short-hard and long-soft bursts than by a model with three components. 

The intention of the GBM catalogs is to provide the community a foundation upon which to perform more
detailed follow-up analysis, taking advantage of the huge dataset of GBM-detected GRBs, and to act as a general reference. Numerous studies using the previous GBM catalogs have been presented elsewhere  \citep[e.g.][]{2014A&A...569A.108K,2015MNRAS.448.2624C,2015AdAst2015E..22P,2015MNRAS.448..403C,2015MNRAS.452..824K,2015ApJ...805L...5A,2015A&A...581A..29T,2017ApJS..229...31H,2017ApJ...841...89A,2019MNRAS.tmp.2436A}. 
Furthermore we emphasize the relevance of the GBM data for multi-messenger astrophysics, which has assumed greater importance following the first
coincident detection of gravitational waves (GW) and electromagnetic (EM) radiation from the same event, namely the  binary neutron star merger event detected by \Fermi-GBM and LIGO on 2017 August 17 \citep{2017ApJ...848L..13A,2017ApJ...848L..14G}. 
Following this ground-breaking discovery, a search of the GBM data for GRBs  with characteristics similar to GRB 170817A was conducted for the full time period of the current catalog \citep{2019ApJ...876...89V}. A total of 13 candidates were identified during 10 mission years, from which it is predicted that \Fermi-GBM will trigger on board on about one burst similar to GRB 170817A per year.

In order to highlight the successful operation of \Fermi-GBM in its first decade of operation, Section \ref{sec:trig-stat} summarizes the 10 year trigger statistics.
The GBM instrument, its data products, and onboard triggering capabilities were discussed in detail in the instrument paper \citep{2009ApJ...702..791M} and previous catalogs. Here we provide a short recap in the introduction of Section \ref{sec:FermiGBM} and in Section \ref{sec:GBMTrigData}. 
Section \ref{sec:InstConfHist} presents the instrument configuration history of the latest four years, which augments the information provided in previous catalogs.
Section \ref{sec:roboBA} introduces a new tool for advanced ground processing, enabled early in 2016, which has been shown to facilitate the daily burst advocate (BA) work. The types of official GBM GCN products (circulars and notices) routinely derived from trigger data are described in Section \ref{sec:GBMNoticeCirculars}. The standard catalog tables are presented in Section \ref{sec:GRBcatAnalysis} and discussed in Section \ref{sec:discussion}. Finally, in Section \ref{sec:Summary}, we conclude with a summary.

\begin{figure}[b]
\begin{center}
\includegraphics[scale=0.6]{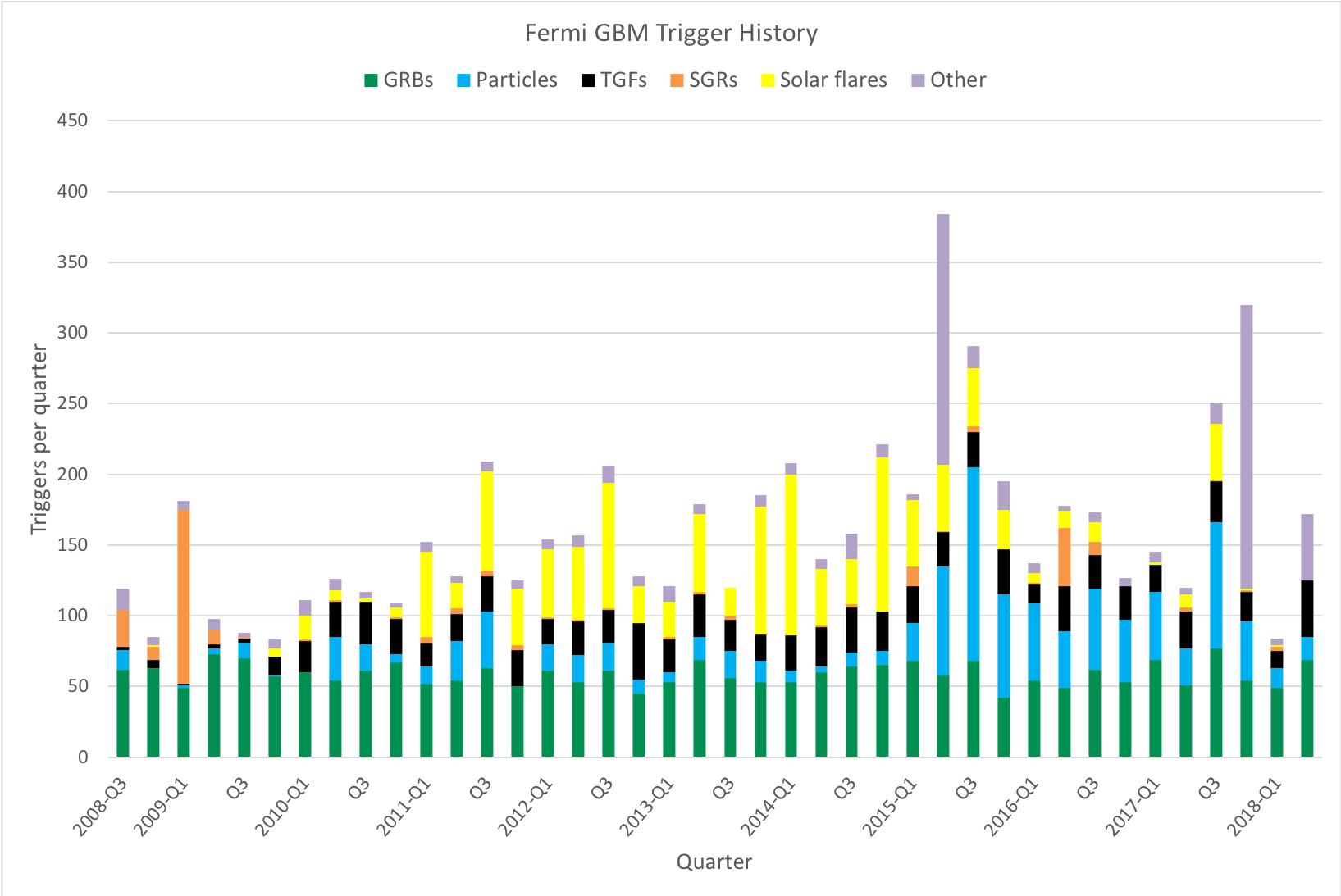}
\caption{\label{fig:quaterlytrigstat} Quarterly trigger statistics over the first 10 years of the mission, starting from 2008 July 12 until 2018 June 30.}
\end{center}
\end{figure}

\section{Trigger Statistics}
\label{sec:trig-stat}

The merit of \Fermi-GBM is best shown by its trigger statistics over the full time range of the current catalog  (see Table \ref{tab:trigstat4thcat} and Figure \ref{fig:quaterlytrigstat}). During its first 10 years of operations, GBM triggered 6434 times, of which 2360 triggers are classified as GRB events. The remaining triggers include events from other cosmic and terrestrial sources, as well as sources of instrumental background such as terrestrial magnetospheric activity.
 
In accordance with the time periods covered by the first three GBM  catalogs, which were two years each, the entries of Table \ref{tab:trigstat4thcat} are subdivided into two year sections. Because the  \nth{4} catalog adds a four year time period, the table lists it in two rows, covering two years each. The last row gives the full 10 year trigger statistics.

Table \ref{tab:trigstat4thcat} lists, in addition to the GRB triggers, the numbers of triggers caused by other sources such as bursts of soft gamma repeaters (SGRs) due to magnetar activity,  triggers on  terrestrial gamma-ray flashes (TGFs), which are connected to thunderstorm activity in the Earth's atmosphere,  and triggers on solar flares (SFs).
Finally, in 2015 and 2017, non-SGR bursting activity from Galactic sources triggered GBM numerous times, as reflected in the column \textquote{Galactic}.

In addition to bursts of gamma-rays, GBM triggers on charged particles (CPs) interacting with the sensitive detector volume, which  are typically magnetospheric events, or, more rarely, cosmic-ray showers. Magnetospheric events occur predominantly in trapped particle regions traversed in the course of  \Fermi's orbit,  mostly in the entry or exit regions of the  South Atlantic Anomaly (SAA) or at high geomagnetic latitude. Rarely do accidental triggers  happen due to background fluctuations or the observed flux is too weak for the trigger source to be identified.  These extra triggers are summed in the \textquote{Other} column of Table \ref{tab:trigstat4thcat}.

Table \ref{tab:trigstat4thcat} also lists the number of Autonomous Repoint Requests (ARRs) generated by GBM during these intervals. The ARR capability allows GBM to repoint the spacecraft in response to particularly bright triggers, thereby bringing the burst direction into the LAT FOV and/or keeping it in the LAT FOV for an extended interval. This capability has been exploited successfully for most of the mission, but it has been disabled since 2018 March 16 due to issues with a stuck spacecraft solar panel.  

The quarterly trigger statistics shown in Figure \ref{fig:quaterlytrigstat} reflect the temporal activity variation of the different kind of sources. The increased trigger rate on solar flare events during the solar maximum period between 2011 and 2017 is obvious, as is the  prolific bursting activity of several magnetars. The latter mostly coincided during the first mission year, during which the activity of three sources predominated: SGR J1550--5418 \citep{2012ApJ...749..122V, 2012ApJ...755..150V}, SGR J0501+4516 and 1E 1841--045. A dedicated catalog summarizes the results on magnetars as observed by \Fermi-GBM in the first five mission years \citep{2015ApJS..218...11C}. 
Non-SGR bursting activity of a few Galactic sources clearly stands out among the \textquote{Other} sources bar in the plot. These are mainly due to the bright source V404 Cyg, a black hole binary, in 2015 (2015-Q2) \citep{2016ApJ...826...37J} and to  Swift J0243.6+6124 in 2017 (2017-Q4), a newly discovered Galactic Be/X-ray binary \citep{2018ApJ...863....9W}. 
Other triggers contributing to the \textquote{Other} sources bar are accidental triggers and triggers with uncertain source classification. A large fraction of the accidental triggers result from the algorithms that use BGO data. The significance levels of these triggers are purposely set low in order to increase the sensitivity for TGFs, which have typical durations much less than the minimum resolution (16 ms) of the data used for triggering.

An increased number of triggers on particle events is observed for the years 2015--2017, mainly by triggers during SAA entry and exit. This could be explained by expansion of the SAA beyond the predefined region stored in the GBM FSW, within which the high voltages of the GBM photo-mulipliers are switched off
and no science data are taken, thereby disabling triggering. 
As expected, the quarterly rate of GRB triggers does not change significantly, fluctuating around a value of 60 triggers/quarter. The rate of triggered TGFs increased in 2009 November (2009-Q4) thanks to an update of the FSW,  improving the capabilities for onboard triggering on TGFs. The actual catalog of \Fermi-GBM TGFs \citep{2018JGRA..123.4381R} includes, in addition to the offline identified TGFs, a list of 686 brighter TGFs, which were able to trigger the GBM FSW, detected since launch in 2008 July 11 through 2016 July 31. This catalog is accessible online\footnote{\Fermi-GBM TGF catalog at FSSC: \url{https://fermi.gsfc.nasa.gov/ssc/data/access/gbm/tgf/}}.

\section{FERMI-GBM: Instrument Overview and Updates}
\label{sec:FermiGBM}

\Fermi-GBM is one of two instruments on the \textit{Fermi Gamma-ray Space Telescope} which was launched on June 11, 2008. GBM is made up of two types of scintillation detectors: 12 NaI(Tl) detectors, sensitive from 8 keV to $\sim$1 MeV, and two BGO detectors, sensitive from 200 keV to 40 MeV. The NaI(Tl) detectors are arranged in four groups of three on the corners of the spacecraft so that they view the whole unocculted sky. The BGO detectors are located on opposite sides of the spacecraft to enable an all-sky view. A detailed description of the instrument, its detectors and electronics can be found in the GBM instrument paper \citep{2009ApJ...702..791M}. 

\subsection{GBM Onboard Triggers and Data Products}
\label{sec:GBMTrigData}

The GBM FSW continuously monitors the counting rates 
in each of several preset energy ranges and timescales and initiates a burst trigger when the rates in two or more detectors exceed fixed thresholds, defined in units of the standard deviation of the background rates. 
A detailed list of the current trigger levels is provided in Section \ref{sec:InstConfHist}.
GBM produces triggered and continuous data types. Triggered data types, available since launch, include accelerated CTIME data (binned to 64 ms, 8 energy channels) and accelerated CSPEC data (binned to 1.024 s, 128 energy channels) for 10 minutes and Time Tagged Event data (individual events at 2 $\mu$s resolution, 128 energy channels) for 5 minutes after a trigger. The continuous data types are CTIME (256 ms, 8 energy channels) and CSPEC (4.096 s, 128 energy channels) available since launch and Continuous Time Tagged Event (CTTE) data (2 $\mu$s, 128 energy channels, available since an FSW update in 2012 November). 

\subsection{Instrument Configuration History}
\label{sec:InstConfHist}

A total of 120 different trigger algorithms may be defined and operated concurrently, each with a specific combination of energy range, timescale, and threshold. Individual trigger algorithms may be disabled or enabled by telecommand. Originally, only data from the NaI(Tl) detectors could be used for triggering. However, beginning in 2009 the FSW was modified to include four additional trigger types that include data from the BGO detectors in their algorithms.
Since launch the available energy ranges for triggering have not been changed. For the NaI(Tl) detectors these are, in units of keV: 25--50, 50--300, $>100$ and $> 300$. For the BGO triggers the energy range from 2 to 40 MeV is used.    
The available trigger timescales range from 0.016 s to 8.192 s in steps of a factor of two. Except for the 0.016~s timescale, pairs of triggers on the same timescale may be offset by half of the time bin to improve the sensitivity \citep{2002ApJ...578..806B}.
The first three GBM GRB catalog papers include the history of enabled triggers and their settings through the first six mission years\footnote{We have to note that the threshold values listed in Table 2 of \citep{2016ApJS..223...28N} are wrong.}. Here we summarize the settings during the subsequent four mission years. Table \ref{trigger:criteria} lists the enabled trigger algorithms at the start of mission year 7, along with their threshold settings, which were not altered subsequently.

\startlongtable
\begin{deluxetable}{ccccccccccccc}
\tabletypesize{\small}
\tablewidth{614pt}
\tablecaption{\label{trigger:criteria} Trigger algorithms at the start of mission year 7}
\tablehead{\colhead{Algorithm} & \colhead{Timescale} & \colhead{Offset} & 
\colhead{Channels} & \colhead{Energy} & \multicolumn{1}{c}{Threshold ($0.1 \sigma$)}\\
\colhead{Number} & \colhead{(ms)} & \colhead{(ms)} & \colhead{} & \colhead{(keV)} & \colhead{2014, July 12}}

\startdata
1 & 16 & 0 & 3--4 & 50--300 & 75 \\
2 & 32 & 0 & 3--4 & 50--300 & 75 \\
3 & 32 & 16 & 3--4 & 50--300 & 75 \\
4 & 64 & 0 & 3--4 & 50--300 & 50 \\
5 & 64 & 32 & 3--4 & 50--300 &  50 \\
6 & 128 & 0 & 3--4 & 50--300 &  50 \\
7 & 128 & 64 & 3--4 & 50--300 &  50 \\
8 & 256 & 0 & 3--4 & 50--300 &  45 \\
9 & 256 & 128 & 3--4 & 50--300 & 45 \\
10 & 512 & 0 & 3--4 & 50--300 & 45 \\
11 & 512 & 256 & 3--4 & 50--300 &  45 \\
12 & 1024 & 0 & 3--4 & 50--300 &  45 \\
13 & 1024 & 512 & 3--4 & 50--300 & 45 \\
14 & 2048 & 0 & 3--4 & 50--300 & 45 \\
15 & 2048 & 1024 & 3--4 & 50--300 & 45 \\
16 & 4096 & 0 & 3--4 & 50--300 & 45 \\
17 & 4096 & 2048 & 3--4 & 50--300 &  45 \\
18\tablenotemark{*} & 8192 & 0 & 3--4 & 50--300 &  50 \\
19\tablenotemark{*} & 8192 & 4096 & 3--4 & 50--300 &  50 \\
20\tablenotemark{*} & 16384 & 0 & 3--4 & 50--300 &  50 \\
21\tablenotemark{*} & 16384 & 8192 & 3--4 & 50--300 & 50 \\
22 & 16 & 0 & 2--2 & 25--50 &  80 \\
23 & 32 & 0 & 2--2 & 25--50 &  80 \\
24 & 32 & 16 & 2--2 & 25--50 & 80 \\
25 & 64 & 0 & 2--2 & 25--50 &  55 \\
26 & 64 & 32 & 2--2 & 25--50 &55 \\
27\tablenotemark{*} & 128 & 0 & 2--2 & 25--50 &  55 \\
28\tablenotemark{*} & 128 & 64 & 2--2 & 25--50 & 55 \\
29\tablenotemark{*} & 256 & 0 & 2--2 & 25--50 & 55 \\
30\tablenotemark{*} & 256 & 128 & 2--2 & 25--50 &  55 \\
31\tablenotemark{*} & 512 & 0 & 2--2 & 25--50 & 55 \\
32\tablenotemark{*} & 512 & 256 & 2--2 & 25--50 & 55 \\
33\tablenotemark{*} & 1024 & 0 & 2--2 & 25--50 & 55 \\
34\tablenotemark{*} & 1024 & 512 & 2--2 & 25--50 & 55 \\
35\tablenotemark{*} & 2048 & 0 & 2--2 & 25--50 & 55 \\
36\tablenotemark{*} & 2048 & 1024 & 2--2 & 25--50 & 55 \\
37\tablenotemark{*} & 4096 & 0 & 2--2 & 25--50 & 65 \\
38\tablenotemark{*} & 4096 & 2048 & 2--2 & 25--50 &  65 \\
39\tablenotemark{*} & 8192 & 0 & 2--2 & 25--50 &  65  \\
40\tablenotemark{*} & 8192 & 4096 & 2--2 & 25--50 & 65 \\
41\tablenotemark{*} & 16384 & 0 & 2--2 & 25--50 &  65 \\
42\tablenotemark{*} & 16384 & 8192 & 2--2 & 25--50 &  65 \\
43 & 16 & 0 & 5--7 & $> 300$ &  80 \\
44\tablenotemark{*} & 32 & 0 & 5--7 & $> 300$ &  80 \\
45\tablenotemark{*} & 32 & 16 & 5--7 & $> 300$ &  80 \\
46\tablenotemark{*} & 64 & 0 & 5--7 & $> 300$ & 60 \\
47\tablenotemark{*} & 64 & 32 & 5--7 & $> 300$ &  60 \\
48\tablenotemark{*} & 128 & 0 & 5--7 & $> 300$ & 55 \\
49\tablenotemark{*} & 128 & 64 & 5--7 & $> 300$ & 55 \\
50 & 16 & 0 & 4--7 & $> 100$ &  80 \\
51\tablenotemark{*} & 32 & 0 & 4--7 & $> 100$ & 80 \\
52\tablenotemark{*} & 32 & 16 & 4--7 & $> 100$ & 80 \\
53\tablenotemark{*} & 64 & 0 & 4--7 & $> 100$ & 55 \\
54\tablenotemark{*} & 64 & 32 & 4--7 & $> 100$ &  55 \\
55\tablenotemark{*} & 128 & 0 & 4--7 & $> 100$ & 55 \\
56\tablenotemark{*} & 128 & 64 & 4--7 & $> 100$ & 55 \\
57\tablenotemark{*} & 256 & 0 & 4--7 & $> 100$ &  55 \\
58\tablenotemark{*} & 256 & 128 & 4--7 & $> 100$ & 55 \\
59\tablenotemark{*} & 512 & 0 & 4--7 & $> 100$ &  55 \\
60\tablenotemark{*} & 512 & 256 & 4--7 & $> 100$ &  55 \\
61\tablenotemark{*} & 1024 & 0 & 4--7 & $> 100$ & 55 \\
62\tablenotemark{*} & 1024 & 512 & 4--7 & $> 100$ & 55 \\
63\tablenotemark{*} & 2048 & 0 & 4--7 & $> 100$ &55 \\
64\tablenotemark{*} & 2048 & 1024 & 4--7 & $> 100$ & 55 \\
65\tablenotemark{*} & 4096 & 0 & 4--7 & $> 100$ &65 \\
66\tablenotemark{*} & 4096 & 2048 & 4--7 & $> 100$ &  65 \\
116\tablenotemark{a} & 16 & 0  & 5--7 & $> 300$  &  55\\
116\tablenotemark{a} & 16 & 0 & BGO/3--6 & 2 - 40 MeV &  55  \\
117\tablenotemark{a} & 16 & 0  & 5--7 & $> 300$ & 45  \\
117\tablenotemark{a} & 16 & 0 & BGO/3--6 & 2 - 40 MeV &  45 \\
118\tablenotemark{a} & 16 & 0 & 5--7 & $> 300$ &  45 \\
118\tablenotemark{a} & 16 & 0 & BGO/3--6 &  2 - 40 MeV & 45 \\
119\tablenotemark{a} & 16 & 0 & BGO/3--6 &  2 - 40 MeV  &  47\\
\enddata

\tablenotetext{*}{Those algorithms have been disabled during most of the mission.}
\tablenotetext{a}{Trigger algorithms using the BGO detector count rates. Algorithm 116 triggers when at least two NaI detectors and  one BGO detector exceed the trigger threshold.  Algorithm 117 is the same as 116, but imposes the additional requirement that the triggered detectors are on the +X side of the spacecraft.  Algorithm 118 is the same as 117, but requires the triggered detectors to be on the -X side of the spacecraft. Algorithm 119 requires a significant rate increase in both BGO detectors independently of the NaI detectors.}

\end{deluxetable}

The low-level threshold (LLT) values are adjustable by telecommand but are generally set at the same values for long periods of time, except for intervals of solar activity when an excessive rate of non-GRB triggers is likely\footnote{A table summarizing the intervals of the non-nominal trigger settings is posted at: \url{https://fermi.gsfc.nasa.gov/ssc/data/access/gbm/llt_settings.html}}.
Since 2012 no modification of the LLT settings has been needed because other flight software settings were used to minimize triggering by the same transient event.
%
The practice of regularly disabling certain soft-energy triggers on weekends and US public holidays, which began in 2012 July, was continued during years 7--10. During weekend times trigger algorithms 22--26 were disabled starting from Friday 15--20 hr UT until Monday 13--20 hr UT  for durations anywhere between 60 and 80 hr. 
Table \ref{tab:trig_mod_hist} includes the trigger algorithm changes during years 7--10, except for the weekend disabling. In the interest of brevity the latter changes are listed separately online.
During mission years 7--10 solar activity continued to be a significant complicating factor affecting the GBM science data. 
In particular, the continuous TTE (CTTE) data mode, which was
implemented beginning in late 2012, may be interrupted or modified 
to mitigate excessive rates of CTTE data, usually caused by solar
activity. In its normal operation this "throttling" uses FSW monitoring of a
variety of data rates to interrupt CTTE data production from the Sun-facing NaI detectors (n0-n5). 
When the more restrictive "aggressive" throttling is enabled, CTTE data production is interrupted
from all 12 NaI detectors. The list below includes the periods when aggressive throttling was
enabled but not the times when throttling actually occurred.
Configuration changes that altered the volume and/or contents of CTTE data are also listed in Table \ref{tab:trig_mod_hist}. 
Listed below are the major configuration changes during mission years 7--10, which are also included in 
Table \ref{tab:trig_mod_hist}.

\begin{description}

\item[8/12/2014] The GBM onboard clock, which counts the elapsed time since 2001 January 1, experienced at 00:38:46 UT a rollover. To minimize unexpected issues with this expected occurrence, all triggers were disabled prior to the rollover.
However, an unexpected impact of the rollover was a high rate of spurious triggering. All triggering was again disabled for a longer period while the issue was studied and the problem (a stale background rates buffer) was corrected.

\item[9/11-15/2014] A further impact of the clock rollover was discovered on 9/11, when a bright solar flare produced a high rate of CTTE data that should have been throttled by the FSW. CTTE mode was disabled while the issue was studied. The cause turned out to be another stale time buffer. The problem was corrected on 9/15 by restarting continuous CTTE mode.

\item[10/22-28/2014] Invoked aggressive throttling of CTTE data.

\item[2/23-26/2015] CTTE data mode was disabled and CTIME data accumulation set to 64 ms, due to flaring activity of SGR 1935+2154.

\item[3/12-13/2015] Long-soft trigger algorithms 25--26 were disabled; algorithms 22--24 were kept enabled.

\item[9/29-10/5/2015] Disabled all TTE data production and set CTIME data accumulations to 64 ms due to elevated solar activity. Also disabled algorithms 25--26 while keeping 22--24 enabled.

\item[3/3/2016] BGO PMT Gain Balance Test was performed. ARRs were disabled; trigger algorithms 116-119 were disabled.

\item[9/6-11/2017] Invoked  aggressive throttling of CTTE data and disabled soft triggers.

\item[11/6-27/2017] Disabled/enabled trigger algorithms 8--17.

\item[3/16-28/2018] GBM was put in safe mode for 12 days due to a spacecraft solar panel drive anomaly\footnote{One of the solar panels stuck and remained in a fixed position at least through the reporting period of this catalog. As a consequence  a modified Fermi rocking strategy has been adopted. A further impact is that ARRs have been disabled since then.}.

\end{description}

\startlongtable
\begin{deluxetable}{rrl}
\tabletypesize{\small}
\tablewidth{360pt}
\tablecaption{\label{tab:trig_mod_hist} Trigger Modification History}
\tablehead{\colhead{Date} & \colhead{Year/DOY/UT} & \colhead{\centering Operation}}
\startdata
8/12/14  &  2014/224:00:27:46   &  Disable triggers   \\
  &  2014/224:00:40:00   &  Enable triggers  \\
  &  2014/224:04:53:46  &  Disable triggers  \\
  &  2014/224:16:29:42  &  Enable triggers  \\
9/11/14  &  2014/254:13:19:15   &  Disable Continuous TTE data  \\
9/15/14  &  2014/258:18:25:22   &  Re-enable Continuous TTE data  \\
10/22/14  &  2014/295:13:48:59    &  Start aggressive throttling of TTE data  and   Disable soft triggers  \\
10/28/14  &  2014/301:19:50:27    &  Set TTE throttling back to normal levels  \\
2/23/15  &  2015/054:17:51:54  &  Turn off Continuous TTE data because of a flaring SGR  \\
2/23/15  &  2015/054:17:52:24  &  Accelerate CTIME data accumulations intervals to 64 ms  \\
2/26/15  &  2015/057:17:21:38  &  Turn on Continuous TTE data production  \\
2/26/15  &  2015/057:17:22:11  &  Decelerate CTIME accumulations  \\
3/12/15  &  2015/071:14:59:58  &  Disable long (64 ms) soft trigger algorithms while keeping short soft algorithms enabled  \\
9/29/15  &  2015/272:15:48:47  &  Turn off all TTE data production because of elevated Solar activity.\\
9/30/15  &  2015/273:14:41:58  &  Accelerate CTIME data accumulation intervals to 64 ms to study SGR's \\
10/2/15  &  2015/275:19:29:03   &  Disable 64 ms soft trigger algorithms 25 and 26 while keeping short soft algorithms enabled \\
10/5/15  &  2015/278:15:24:22  &  Turn on all TTE data production  \\
10/5/15  &  2015/278:18:39:02  &  Decelerate CTIME accumulations  \\
3/3/16  &  2016/063:12:20:00   &  Disable the BGO AGC's  (Collect data for BGO PMT Gain Balance Test)  \\
  &  2016/063:12:20:05   &  Disable ARRs  \\
  &  2016/063:12:20:10   &  Turn off algorithms 116--119 (TGF algorithms)  \\
  &  2016/063:12:20:15   &  Turn off PMTs 12 and 14 (BGO detectors)  \\
  &  2016/063:18:00:15   &  Turn on PMTs 12 and 14   \\
  &  2016/063:23:50:00   &  Turn on PMTs 13 and 15   \\
  &  2016/063:23:50:05   &  Set ARR threshold back to McIlwain 158  \\
  &  2016/063:23:50:10   &  Enable the BGO AGC's  \\
  &  2016/063:23:50:15   &  Re-enable the TGF trigger algorithms (116--119)  \\
9/6/17  &  2017/249:19:33:33  &  Aggressively throttle TTE data  and   Disable soft triggers  \\
9/7/17  &  2017/250:15:37:34  &  Set throttling of TTE data back to normal; Re-enable the soft trigger algorithms  \\
9/7/17  &  2017/250:19:15:49  &  Aggressively throttle TTE data  and   Disable soft triggers   \\
9/7/17  &  2017/250:21:49:09  &  Set throttling of TTE data back to normal; Re-enable the soft trigger algorithms.  \\
9/8/17  &  2017/251:15:02:41  &  Aggressively throttle TTE data  and   Disable soft triggers   \\
9/11/17  &  2017/254:16:24:20  &  Set throttling of TTE data back to normal; Re-enable the soft trigger algorithms  \\
11/3/17  &  2017/307:22:02:48  &  Disable the algorithms 12--15  \\
11/6/17  &  2017/310:21:19:15  &  Disable algorithms 10--17  \\
11/7/17  &  2017/311:20:06:20  &  Disable algorithms 10--17  \\
11/8/17  &  2017/312:19:05:54  &  Disable algorithms 8--17  and   25--26  \\
11/15/17  &  2017/319:22:52:32  &  Re-enable algorithms 8 and 9  \\
11/21/17  &  2017/325:18:06:51  &  Re-enable algorithms 8 through 11  \\
11/27/17  &  2017/331:19:22:30  &  Re-enable algorithms 12 through 17 \\
3/16/18 & 2018/075:05:12:00    &  GBM Turned to safe mode  \\
3/28/18  &  2018/087:13:43:55  &  Boot GBM after safe mode  \\
  &  2018/087:13:59:31  &  HV on  \\
  &  2018/087:14:04:14  &  TTE on  \\
  &  2018/087:14:05:56  &  Master start to enable triggers \\
\enddata
\end{deluxetable}

\subsection{Advanced Ground Processing}
\label{sec:roboBA}
Beginning in early 2016, an automated localization algorithm, called the RoboBA, was placed into operation within the GBM Burst Alert Pipeline (BAP). The RoboBA is a set of automated algorithms developed to replace the Human-in-the-Loop (HitL) localization for most GRB triggers.  HitL processing requires BAs to be on call at all times and ready to promptly localize the GBM low-latency trigger data, which has faced a median 1--2 hr latency.  Due to the increasing interest in and importance of  GBM-detected and localized GRBs, localizations of GRBs are desirable as soon as possible.  The RoboBA now provides localizations for GRBs with accuracy equivalent to the human BA processing for GRBs within 10 minutes after trigger.  The RoboBA also provides a preliminary estimate in the GCN notice whether the GRB is likely to be a short or long duration GRB, with a success rate of $>85$\% when compared to the final $T_{90}$.  Once the RoboBA performs the localization, the BAP submits a GCN Notice and Circular for the RoboBA localization. The pipeline automatically creates the localization products, including the full-sky HEALPix map of the localization incorporating the estimated systematic uncertainty model, and uploads them to the HEASARC. The RoboBA has a complete end-to-end automatic processing success rate of $> 80$\%, with most failures due to dropped data packets in the real-time communication stream from the spacecraft.  RoboBA catches these failures and reports them to the human BA so that a manual localization can be performed.  A detailed description of the RoboBA algorithm, an evaluation of its effectiveness, and the improvements implemented in the algorithm in 2019 are described in~\citet{2019arXiv190903006G}.

\subsection{FERMI-GBM GCN Notices and Circulars}
\label{sec:GBMNoticeCirculars}

Here we list the actual Gamma-ray burst Coordinates Network (GCN) notices and circulars relevant for GRBs as announced in 2019 \citep{2019GCN.24408....1F}, which are  released automatically or by the GBM BA.

\subsubsection*{Official Fermi-GBM notices:}

Notices are automated, standard format text messages designed to be easily parsed by a computer. They are typically low latency, within tens of seconds to 10 minutes of the GRB trigger, and can be found here \url{https://gcn.gsfc.nasa.gov/fermi_grbs.html} or subscribed to through GCN.

\begin{itemize}
    \item GBM Alert: Initial alert includes trigger time, trigger significance, trigger algorithm, trigger timescale.
    \item GBM Flight position: Fermi-GBM onboard calculated localization, generated onboard Fermi, tens of seconds after trigger (may be multiple notices).
    \item GBM Ground position: Intermediate ground localization based on latest onboard generated GRB detector rate information (may be multiple notices).
    \item GBM Final Position: reports the ground generated RoboBA or Human-in-the-Loop (BA) final localization and whether the GRB was likely long or short; 10 minutes after trigger using the full trigger dataset.
    \item Fermi-GBM SubThreshold: reports the time, duration, localization, and reliability for candidate short GRBs found in ground searches of CTTE data (latency here is longer than 10 minutes). See \url{https://gcn.gsfc.nasa.gov/fermi_gbm_subthreshold.html}
\end{itemize}

\subsubsection*{Official Fermi GBM circulars:}

Circulars are reports of follow-up observations made by the observers.

\begin{itemize}
    \item GRB YYMMDDX: Fermi GBM Final Real-time Localization. Introduced in 2019 July. Automated circulars reporting the final real-time localization and HEALPix map based on our automatic processing (RoboBA, operational since early 2016). These circulars are issued for all GRBs that RoboBA localizes and include initial information about the burst duration (likely SHORT/LONG).
    \item GRB YYMMDDX: Fermi GBM Detection (RoboBA or Human in the Loop generated Final localization, spectral analysis, and burst duration)
    \item GRB YYMMDDX: Fermi GBM Observation (spectral analysis, and burst duration for GRBs better localized by another instrument)
\end{itemize}

\section{GRB Catalog Tables}
\label{sec:GRBcatAnalysis}
Here we present the standard catalog tables, listing in Table \ref{tab:main_table} all 2360 triggers 
of the first decade of GBM operation that were classified as GRBs\footnote{The total number of GBM-detected GRBs is four less, since GBM triggered twice on each of four GRBs}. The associated catalog analysis results for each trigger are shown in Table \ref{tab:durations} for the duration analysis and in Tables  \ref{tab:pf_fluence} and \ref{tab:pf_fluence_b} for the peak flux and fluence  analysis  in two energy ranges. The GRB catalog compilation and analysis process has not changed since the production of the latest GRB catalog, and is described in detail in previous catalog papers. 
The standard tables of the newest catalogs always include the GRB entries of the previous catalogs, with only some minor updates for some individual GRBs, where a reanalysis was necessary. 
There are two browsable catalogs accessible online at HEASARC, FERMIGTRIG\footnote{\Fermi-GBM trigger catalog at HEASARC: \url{https://heasarc.gsfc.nasa.gov/W3Browse/fermi/fermigtrig.html}} and FERMIGBRST\footnote{\Fermi-GBM burst catalog at HEASARC: \url{https://heasarc.gsfc.nasa.gov/W3Browse/fermi/fermigbrst.html}}. All GBM triggers are entered in FERMIGTRIG, but only those triggers classified as bursts are entered in the FERMIGBRST catalog. Thus, a burst will be listed twice, once in FERMIGTRIG and once in FERMIGBRST. The burst catalog analysis requires human intervention; therefore, GRBs will be entered in the trigger catalog before the burst catalog. The latency requirements are one day for triggers and three days for bursts. 

\subsection{GRB  Localizations  and Trigger Characteristics}

\begin{figure}[t!]
	\begin{center}
			\includegraphics[scale=0.3]{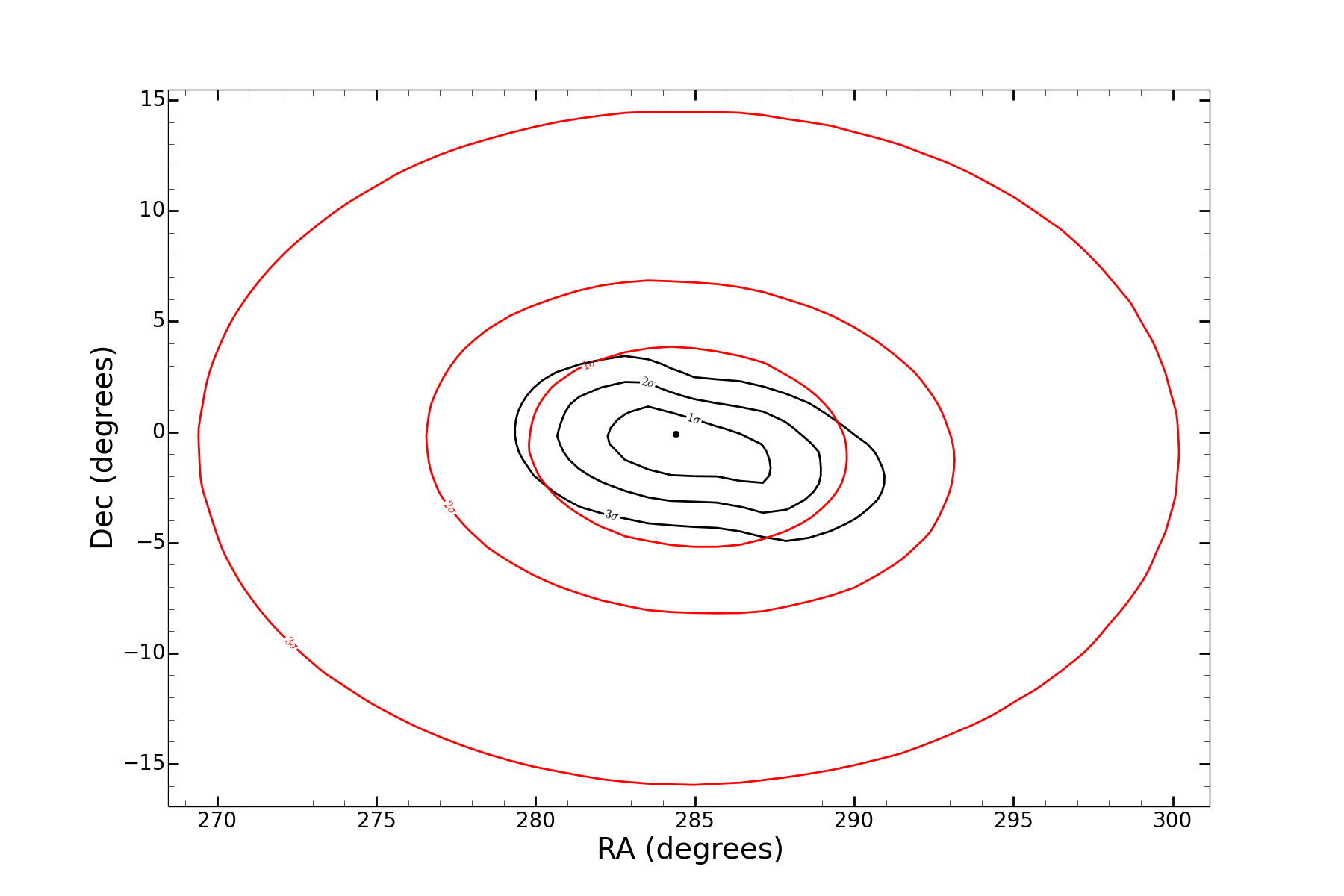}
	\end{center}
\caption{Probability map showing the statistical (black contours) and systematic (red contours) uncertainties of the GRB 170208C localization.
\label{fig:ProbMap}}
\end{figure}

The catalog analysis is based on using the most reliable source locations for the determination of the instrument response (Detecor Response Matrix; DRM). This is quite important since all of the analysis results depend on the response files generated for the particular GRB location. 
These locations are listed in Table \ref{tab:main_table} and are adopted from the BA (HitL) and RoboBA analysis results, which were uploaded to the GBM trigger catalog at the GIOC (with a copy at HEASARC\footnote{\url{https://heasarc.gsfc.nasa.gov/FTP/fermi/data/gbm/bursts/}}). 
The GBM location uncertainties shown in the table are the circular area equivalent of the statistical uncertainty (68\% confidence level).
There is additionally a systematic error that we have characterized for HitL localizations as a core-plus-tail model, with 90\% of GRBs having a 3.7 deg error and a small tail suffering a larger than 10 deg systematic error \citep{2015ApJS..216...32C}. 
An evaluation of automated \Fermi-GBM localizations is presented in \cite{2019arXiv190903006G}, showing that the latest version of RoboBA  yields significant improvement in the systematic uncertainty, removing the long tail identified in the systematic, and improves the overall accuracy. The systematic uncertainty for the updated RoboBA localizations is 1.8 deg for 52\% of GRBs and 4.1 deg for the remaining 48\%.
Probability maps reflecting the total uncertainty on a GBM GRB location, which are the  convolution of the statistical uncertainty with the best current model for the systematic errors have been routinely delivered to the HEASARC since 2014 January, and have also  been processed and delivered for the GRBs prior to 2014. An example localization contour map for GRB 170208C in shown in Figure \ref{fig:ProbMap}. 

Non-GBM locations are listed for bursts that were detected by an instrument providing a better location accuracy such as LAT, the Swift Burst Alert Telescope  \citep[BAT; ][]{2005SSRv..120..143B} or X-ray Telescope \citep[XRT; ][]{2005SSRv..120..165B}, INTEGRAL \citep{2003A&A...411L.291M},  or were localized more precisely by the Inter-Planetary Network  \citep[IPN; ][]{2013ApJS..207...39H}. 
The higher-accuracy location source is listed under the column \enquote{Location Source}, which lists only the name of the mission rather than the specific instrument on board that mission (e.g., Swift implies the locations are either from Swift-BAT or Swift- XRT or Swift-UVOT).
The errors on the GRB locations determined by other instruments are not necessarily 1$\sigma$ values. For the GBM analysis, a location accuracy better than a few tenths of a degree provides no added benefit
because of significant systematic errors in GBM location. 

The first column of Table \ref{tab:main_table} lists the GBM Trigger ID along with a conventional GRB name in the second column as defined by the GRB-observing community\footnote{ 
Note that the entire table is consistent with the small change in the GRB naming convention that became effective on 2010 January 1 \citep{2009GCN.10251....1B}: if for a given date no burst has been “published” previously, then the first burst of the day observed by GBM includes the “A” designation even if it is the only one for that day.}. For year 5--10, only GBM-triggered GRBs for which a  GCN Circular was issued are assigned a GRB name. 

The criterion for issuing a GBM Detection/Observation Circular is if a GRB was either detected by any other mission (as listed in the last column of Table \ref{tab:main_table}) or if it generated an ARR to the Fermi spacecraft or the count rate in the 50--300 keV energy range summed over the triggered detectors exceeded 1000 counts per second above the background. This arbitrary number was chosen at the beginning of the mission to focus on brighter events and not to issue too many circulars. During the 10 year period of the catalog for about \sfrac{1}{3}  of the GRBs  Fermi GBM Detection / Observation Circulars were released.     

The third column lists the trigger time in universal time (UT). 
Table \ref{tab:main_table} also shows which algorithm was triggered, along with its timescale and energy range. Note that the listed algorithm is the first one to exceed its threshold but it may not be the only one. The table also lists other instruments that detected the same GRB\footnote{This information was drawn from the GCN archive accessible at \url{http:// gcn.gsfc.nasa.gov/gcn3_archive.html}. A more complete list of detections is available at \url{http://www.ssl.berkeley.edu/ipn3/masterli.txt}}. Finally, we identify the GBM GRBs for which an ARR was issued by the GBM FSW in the last column of Table \ref{tab:main_table}. 
A total of 172 GRBs (7.3\% of the total) were followed by ARRs during the first  nine years of Fermi, although the spacecraft might not have slewed in every case for technical reasons, such as Earth limb constraints. The majority of these ARRs were due to high peak fluxes. In addition, there were 48 ARRs that were issued for non-GRB triggers because of  misclassification by the GBM FSW.

\begin{figure}[t!]
\begin{center}
\epsscale{1.0}
\plotone{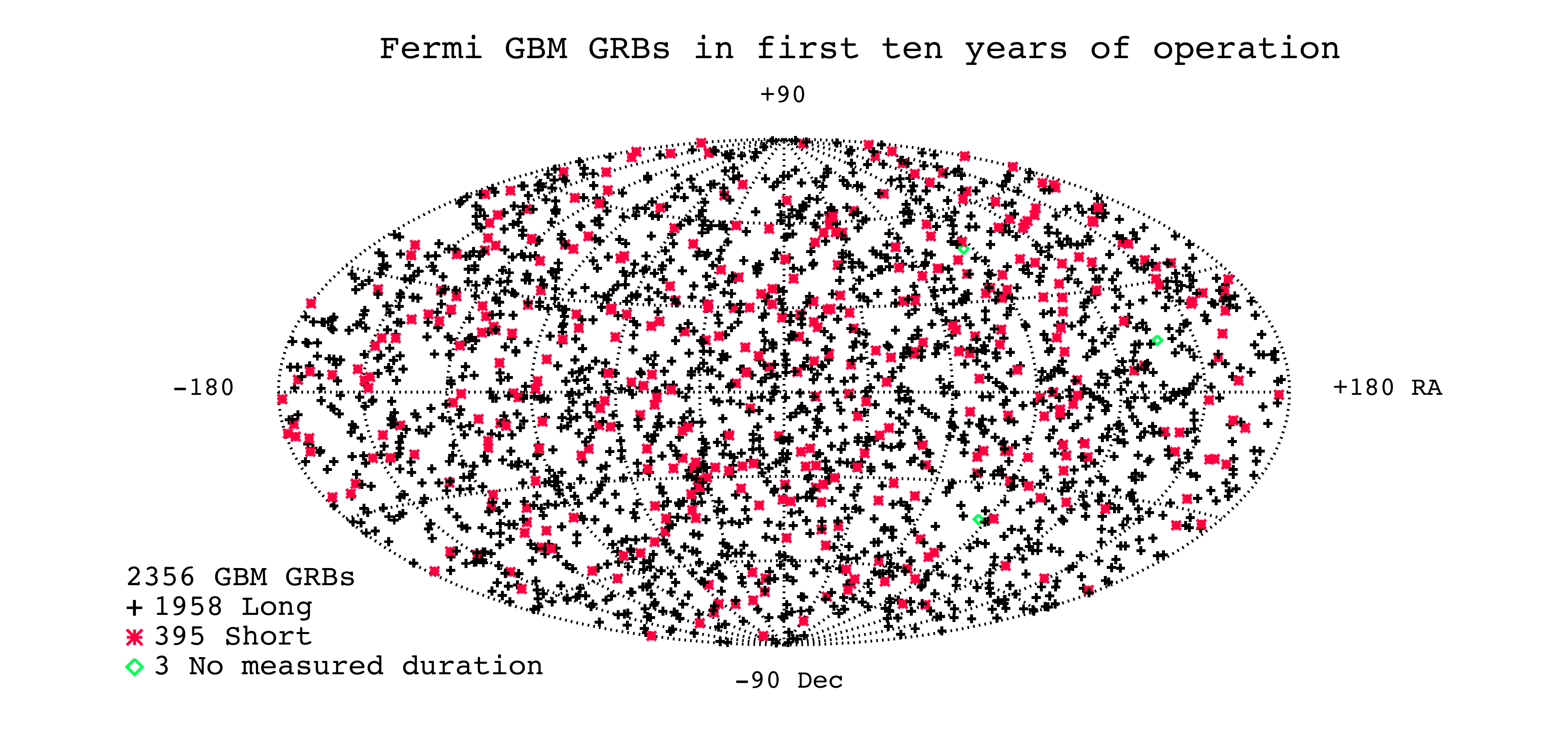}
\caption{\label{fig:sky_dist} Sky distribution of GBM-triggered GRBs in celestial coordinates. Crosses indicate long GRBs ($T_{90} > 2$~s); asterisks indicate short GRBs.}
\end{center}
\end{figure}

\subsection{GRB Duration, Peak Flux, and Fluence}

The analysis performed to derive the duration, peak flux and fluence of each burst (as listed in Tables \ref{tab:durations}--\ref{tab:pf_fluence_b}) is based on an automatic batch fit routine implemented within the RMFIT software\footnote{The spectral analysis package RMfit was originally developed for time- resolved analysis of BATSE GRB data but has been adapted for GBM and other instruments with suitable FITS data formats. The software is available at the Fermi Science Support Center: \url{https://fermi.gsfc.nasa.gov/ssc/data/p7rep/analysis/rmfit/}}. It uses a forward-folding technique to obtain the best-fit parameters for a chosen model given user-selected source and background time intervals in the 10--1000 keV energy range from data files containing observed count rates and a corresponding detector response matrix. 
Burst durations \tf~ (\tn) are determined from the interval between the times where the burst has reached 25\% (5\%) and 75\% (95\%) of its maximum fluence. The burst durations \tf~ and \tn~ listed in Table \ref{tab:durations} were computed in the 50--300 keV energy range. This is primarily due to the fact that GRBs have their maximum spectral density in this energy range. In addition, this energy range makes it easier to compare the present results with those of the predecessor BATSE. 
In addition the table provides the respective 1$\sigma$ error estimates \citep{1996ApJ...463..570K} and start times relative to the trigger time. For a few GRBs, the duration analysis could not be performed either because the event was too weak or due to technical problems with the input data. Also, it may be noted that the duration estimates are only valid for the portion of the burst that is visible in GBM light curves summed over those NaI(Tl) detectors whose normals make less than $60^\circ$ to the source. If the burst was partially occulted by Earth or had significant emission while GBM detectors were turned off in the SAA region, then the “true” durations may be underestimated 
or are not reliable, depending on the intensity and variability of the undetected burst emission. GRBs that triggered while \Fermi\, was close to SAA or where the trigger is unusual in any other way, are indicated in Tables  \ref{tab:main_table}, \ref{tab:durations} by a footnote. 

For technical reasons, it was not possible to perform a single analysis of the unusually long GRB 091024A \citep{2011A&A...528A..15G}, GRB 130925A, and GRB 150201A, and so the analysis was carried out separately for the two triggered episodes. Similarly, GRB130925A also had three emission episodes well separated in time, for which GBM triggered on the first two episodes. These cases are also noted in the Tables \ref{tab:main_table}, \ref{tab:durations}. The reader may note that for  most GRBs, the present analysis used data binned no finer than 64~ms, and so the duration estimates (but not the errors) are quantized in units of 64~ms. 
However, for 
extremely short events, TTE/CTTE data were binned with widths of 32~ms, 16~ms or even 8~ms in 
about 4\% of the cases, which was necessary in order to resolve the GRBs.

As a part of the duration analysis, peak fluxes and fluences were computed in two different energy ranges. Table \ref{tab:pf_fluence} shows the values in 10--1000 keV and Table \ref{tab:pf_fluence_b} shows the values in 50--300 keV. The analysis results for low fluence events are subject to large systematic errors primarily because they use 8-channel spectral data and should be used with caution. The fluence measurements in the accompanying spectroscopy catalog (S.~Poolakkil et al. 2020, in preparation), which uses the 128-channel CSPEC or TTE data, are more reliable for such weak events.
The peak fluxes for each burst were computed in the same energy ranges and for three different timescales: 64, 256, and 1024 ms. Since only 20\% of the
bursts have detectable emission in the BGO detectors\footnote{ GRBs with significant emission in at least one BGO detector  above 300~keV were highlighted in the main GRB tables of the first two catalogs \citep{2012ApJS..199...18P,2014ApJS..211...13V}. In the first four years there were 204 BGO bright GRBs out of 954 GRBs.}, only NaI detector data were used for the catalog analysis.

\begin{figure}[h!]
    \centering
    \includegraphics[width=0.45\columnwidth]{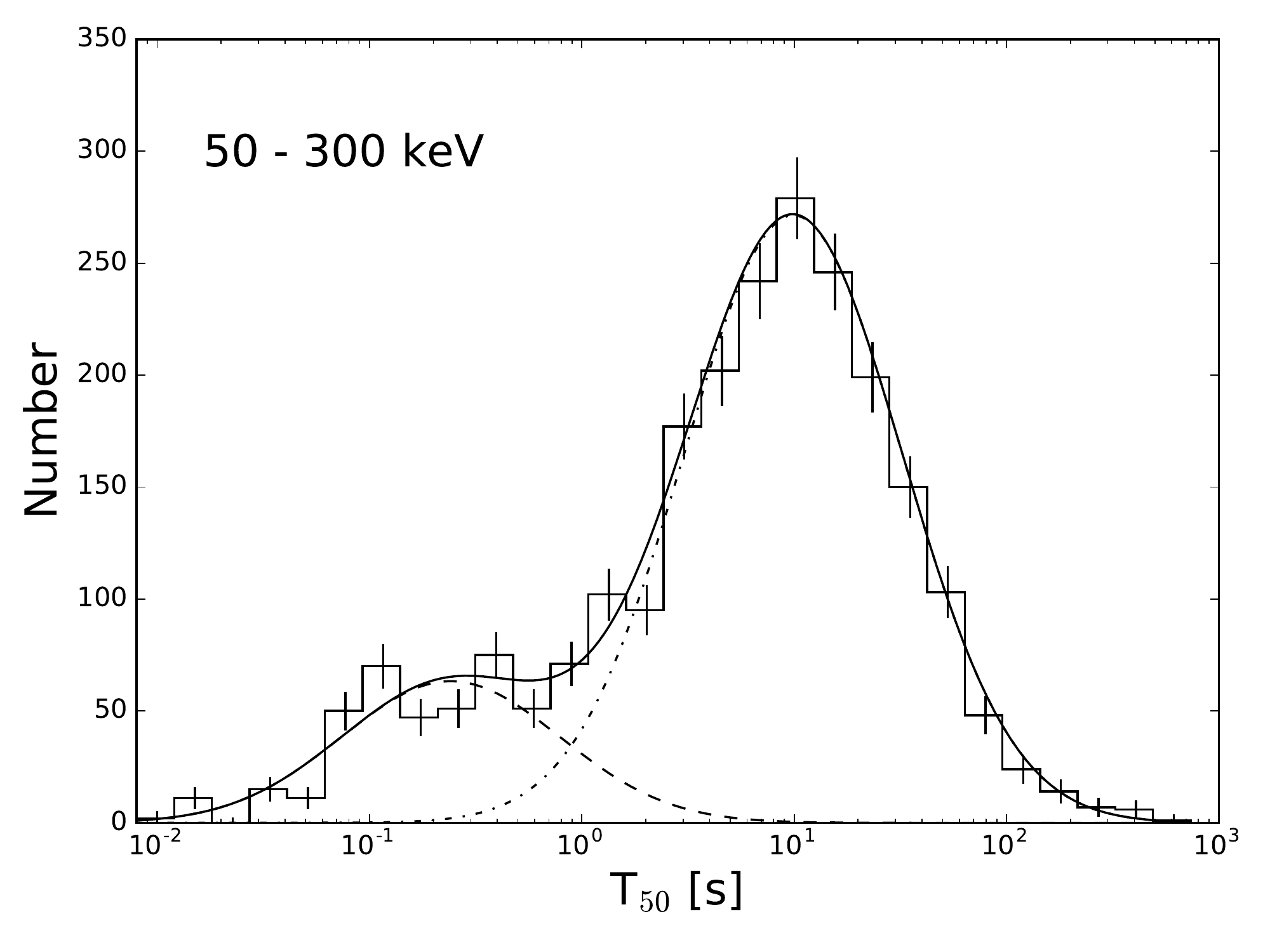}
    \includegraphics[width=0.45\columnwidth]{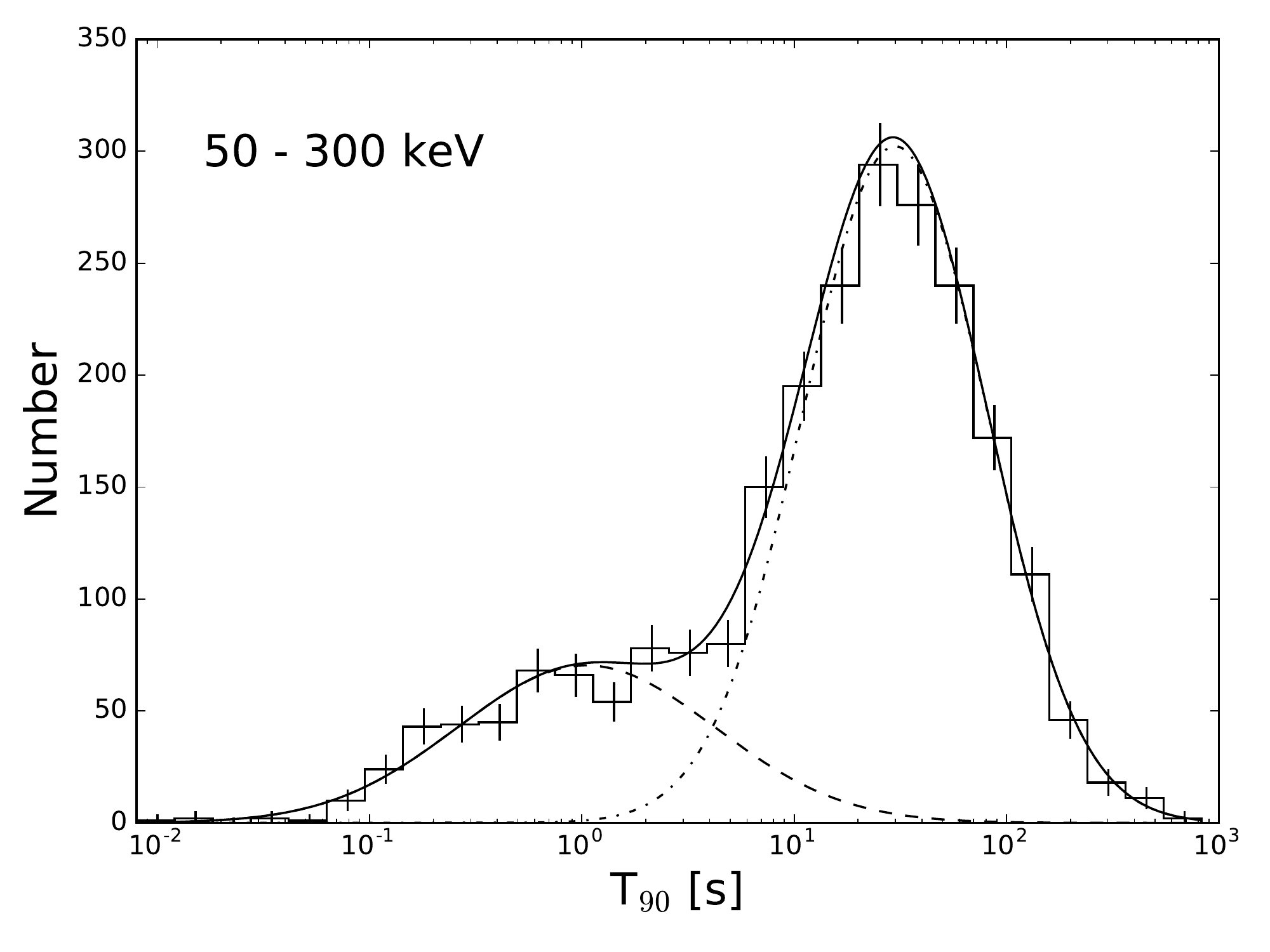}
    \caption{$T_{50}$ (left) and $T_{90}$ (right) distributions. Lines show the best-fitting models.}
    \label{fig:Txxdistr}
\end{figure}

\section{Discussion}
\label{sec:discussion}
Here we provide the standard set of figures as shown in the previous catalogs. The sky distribution of GBM-triggered GRBs in celestial coordinates is shown in Figure \ref{fig:sky_dist}, still reflecting now for the large 10 yr sample that both the long and short GRB locations do not show any obvious anisotropy, which is consistent with an isotropic distribution of GRB arrival directions. 
The histograms of the logarithms of GBM-triggered GRB durations (\tf\, and \tn) are shown in Figure \ref{fig:Txxdistr}. Using the conventional division between the short and long GRB classes (\tn $\le 2$~s and \tn $> 2$~s, respectively), we find that during the first 10 years there were 395 (17\%) short GRBs  and 1958 (83\%) long GRBs\footnote{For 3 GRBs the duration measurement using our standard method was not possible.}. 

We fit the duration distributions using the unbinned, maximum likelihood method, Mclust \citep{Fraley+02mclust}. This method assumes the components have log-normal distribution and decides the optimal number of groups using a Bayesian Information Criterion. We find both of the duration distributions are best described by a two component model corresponding to the short and long GRB categories (see Figure \ref{fig:Txxdistr}), which reaffirms the study of  \cite{2016ApJS..223...28N}
which didn't provide any clue for an extra class, like  soft-intermediate duration GRBs bridging the other two groups. The results of the fits are presented in Table \ref{tab:tdistr},
$N(\mu, \sigma, w)$ is a single Gaussian for the $\log (T_{50})$ or $\log (T_{90})$ distribution, where $\mu$ represents the mean, $\sigma$ is the standard deviation and $w$ is the weight of the component. 

We fit the hardness-duration distributions with the same method, and find that the best-fitting solutions are not meaningful. Namely, the algorithm finds three components: one that can be associated with the short group, and two other groups that divide the long population essentially along a constant duration with approximately equal weights. This structure does not correspond to previous studies that find three groups \citep{2006A&A...447...23H,2010ApJ...725.1955V}. However, it might point to an asymmetric distribution \citep{2019ApJ...870..105T}.
\begin{table}[htb]
    \centering
    \begin{tabular}{c|ccc|ccc}
         &  $\mu$ (short) &  $\sigma$ (short) & $w$ (short) &  $\mu$ (long) & $\sigma$ (long) & $w$ (long) \\
     $T_{50}$    & -0.618 ($\rightarrow 0.241$\,s) &0.265 &0.189 & 0.995 ($\rightarrow 9.89$\,s) & 0.265 & 0.811 \\
     $T_{90}$    & 0.0208 ($\rightarrow 1.05$\,s) &0.367 & 0.245 & 1.476 ($\rightarrow 29.9$\,s)  & 0.189 & 0.755 \\
    \end{tabular}
    \caption{Parameters of the Gaussian distributions for $\log (T_{50})$ and $\log (T_{90})$. We display the actual mean duration in parentheses ($=10^{\mu}$). Note that for the T$_{50}$ distribution the fitting procedure yields a solution where the variances of the two components are equal thus reducing the number of degrees of freedom. }
    \label{tab:tdistr}
\end{table}

Checking in Figure  \ref{fig:Txxdistr} for the \tn\, duration at which the lognormal fits to the short and long GRB classes intersects, indicating a 50\% probability that a GRB is in fact long/short we derive a \tn\, value of 4.2 s. Using this value as division of the short/long GRB populations we get 532 (22.5\%) short and 1821 (77.5\%) long GRBs. 

\begin{figure}[b]
\begin{center}
\epsscale{1.05}
\plottwo{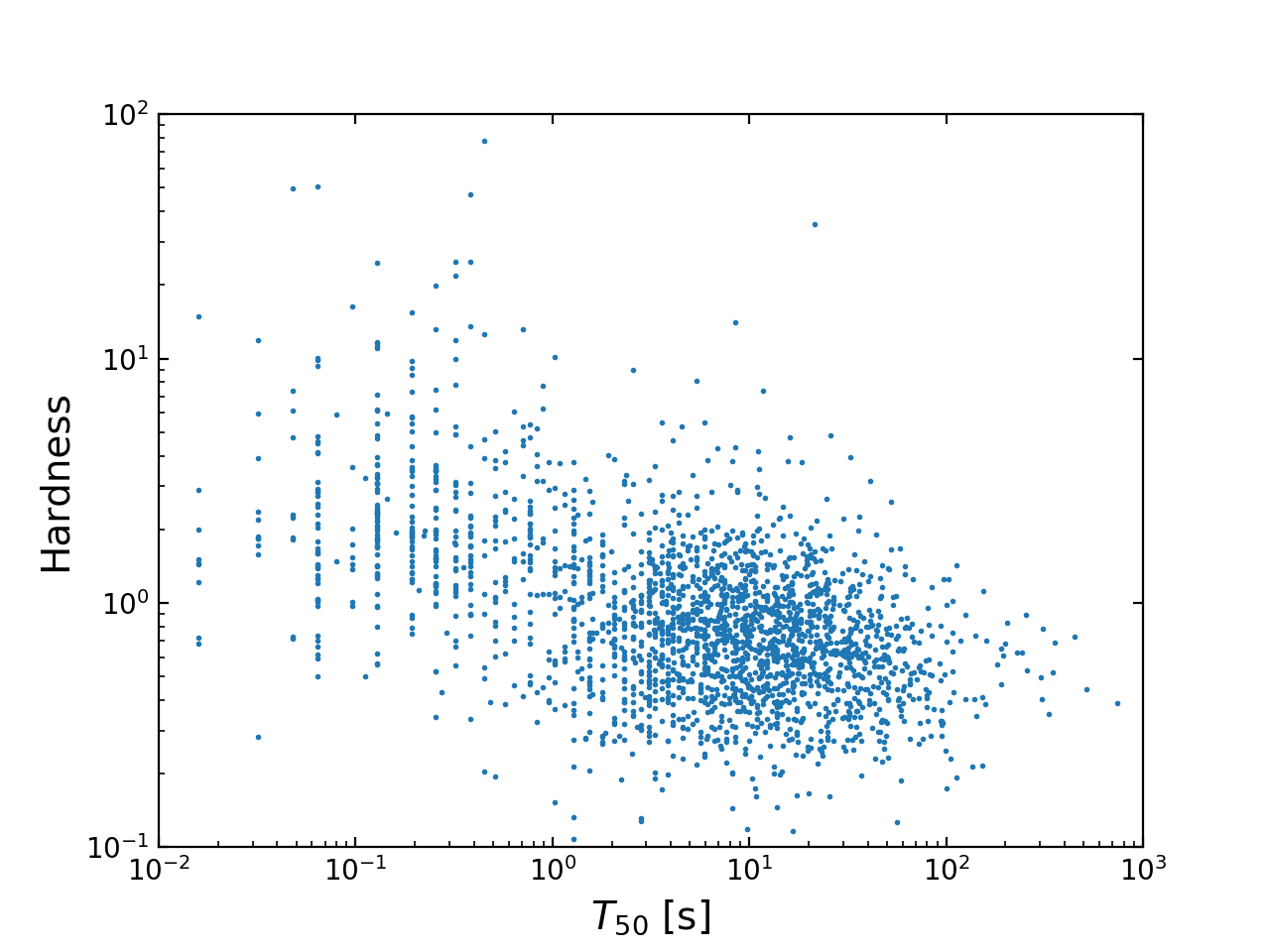}{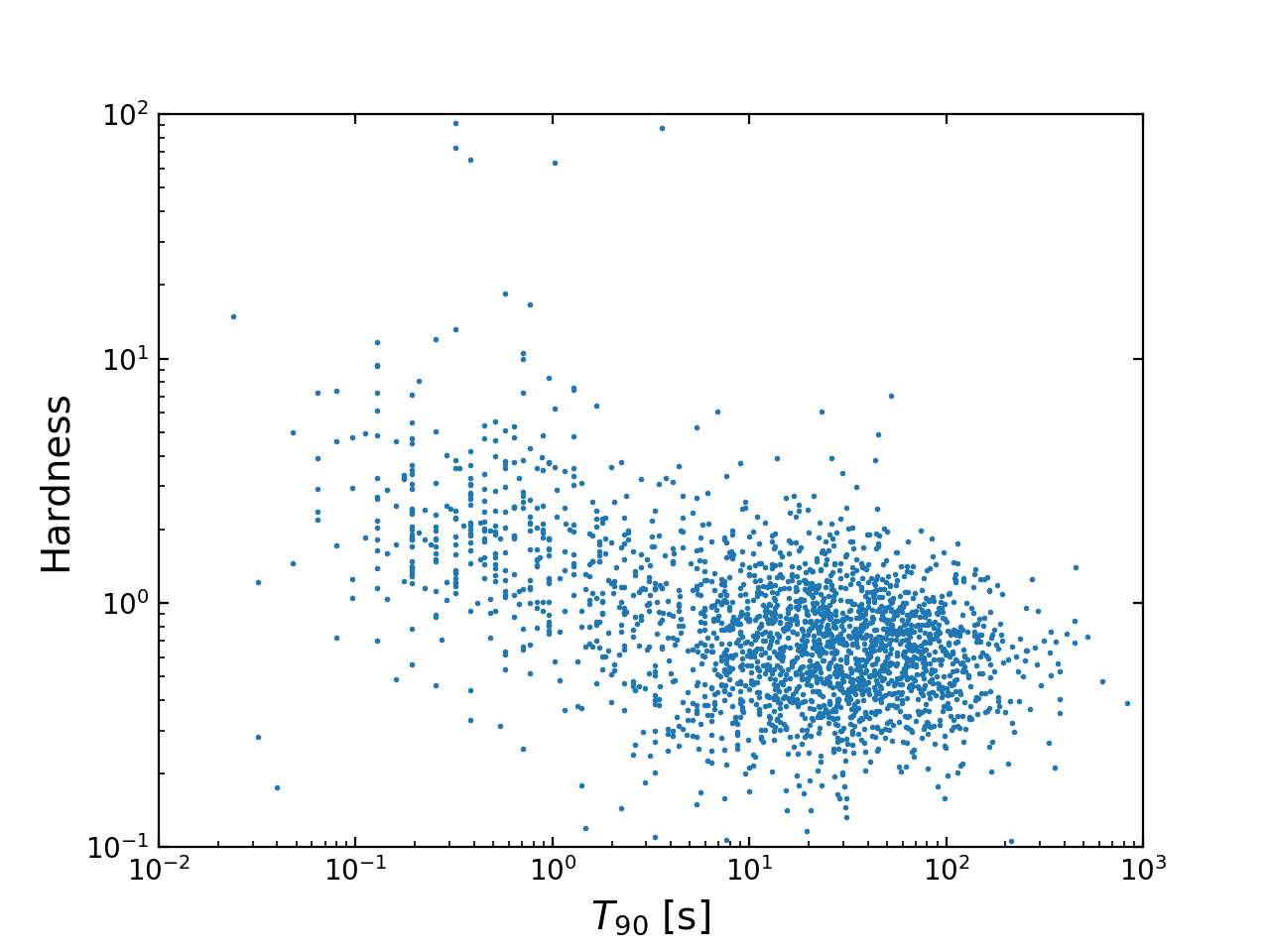}
\caption{\label{fig:hardness_vs_dur} Scatter plots of spectral hardness vs.\ duration are shown for the two duration measures $T_{50}$ (left plot) and $T_{90}$ (right plot). 
The estimated errors for both quantities are not shown but can be quite large for the weak events. Nevertheless, the anti-correlation of spectral hardness with burst duration is evident.}
\end{center}
\end{figure}

Figure \ref{fig:hardness_vs_dur} shows scatter plots of the spectral hardness versus \tf- and \tn- durations.
The spectral hardness is obtained from spectral fits for each GRB, by using the photon model fit parameters, which are a byproduct of the duration analysis. By summing the deconvolved counts in each detector and time bin in two energy bands (10--50~keV and 50--300~keV), and further summing each quantity in time over the $T_{50}$ and $T_{90}$ intervals, the hardness was calculated separately for each detector as the ratio of the flux density in 50--300 keV to that in 10--50 keV, and finally averaged over detectors.

\begin{figure}
\begin{center}
\epsscale{0.7}
\plotone{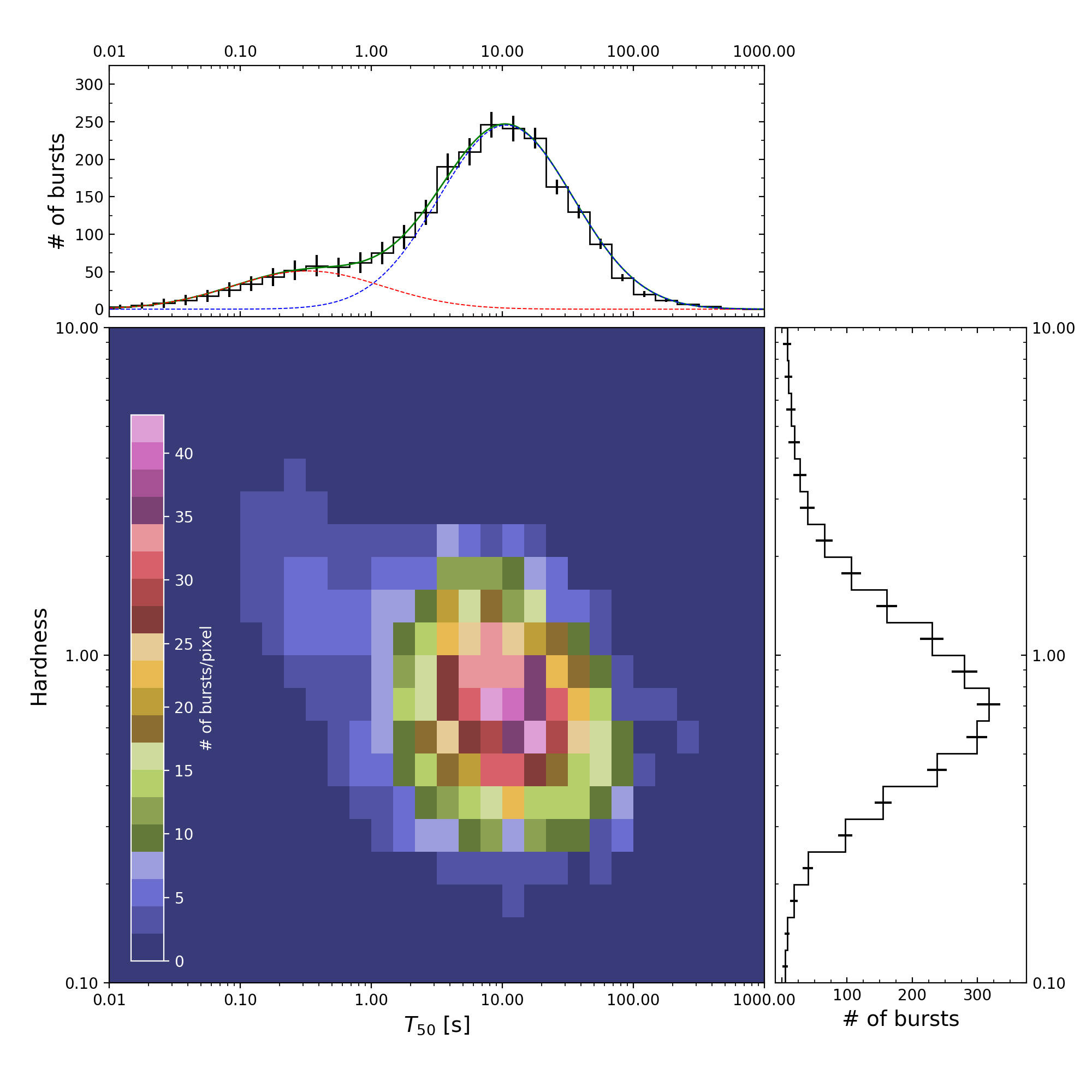}
\plotone{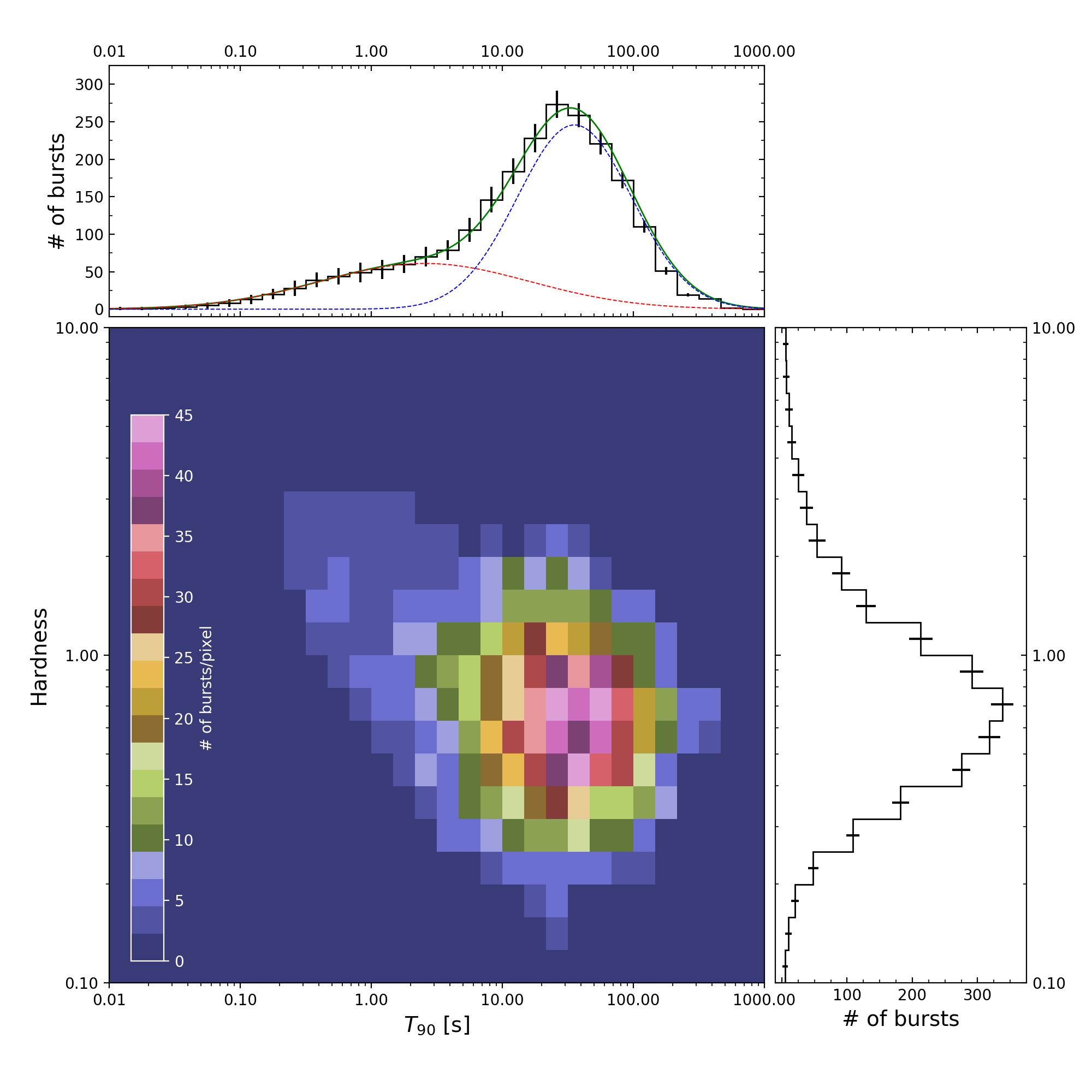}
\caption{\label{fig:hardness_vs_dur_2dhist} Two-dimensional histogrammed probability density plots of the spectral hardness vs. duration (top: \tf, bottom: \tn) accounting for the uncertainties of both parameters. The color bar provides the color mapping for the number of bursts per pixel. The plots attached to the  top and right are the projections of the  individual histogrammed probability densities of duration and hardness.
Lognormal bimodal fits (green line) together with individual lognormal fits to the long (blue line) and short (red line) GRB classes are overplotted in the duration histograms.}
\end{center}
\end{figure}

The estimated errors derived from the duration and hardness analysis are not included in the duration distributions and scatter plots shown in Figures \ref{fig:Txxdistr} and \ref{fig:hardness_vs_dur}.
A more realistic representation of these parameters incorporating the uncertainty is shown in Figure \ref{fig:hardness_vs_dur_2dhist} as a histogrammed probability density plot. It was derived via Monte Carlo sampling from the probability density function (PDF) of the duration and hardness parameters and their standard deviations for each GRB \citep[][S.~Poolakkil et al. 2020, in preparation]{2016ApJ...818...18G}. By randomly selecting a value from each of those PDFs, sorting them into duration/hardness histograms with predefined bins, and additionally into pixels of the corresponding hardness-duration plane, then repeating the procedure for a number of iterations (typically $>$ 1000), we were able to derive PDFs for each histogram bin and map pixel.
We choose the median as the centroid of the frequency value of each bin/pixel and the error bars shown in the duration and hardness histograms represent the 68\% credible interval centered at the median. 

The bimodal shape of the duration distributions shown in Figure \ref{fig:hardness_vs_dur_2dhist} is less distinct
compared to the representation of the duration distributions shown in Figure \ref{fig:Txxdistr}.
Again assigning  GRBs to the short/long GRB classes by using the intersection of the two lognormal fits, we obtain a
value for \tn\, 
of 6.1 s, which is approximately 1.5 times the value derived from Figure  \ref{fig:Txxdistr}, now yielding 615 (26\%) in the short  and 1738 (74\%) in the long  GRB class. 

It emerges that the representation of the duration distributions  as histogrammed probability density plots 
suggests an increased  number of GBM GRBs that could be attributed to the short GRB class. This supports the result of the search for GBM GRBs similar to GRB 170817A, presented in \cite{2019ApJ...876...89V}, which already revealed candidate short GRBs  with a \tn\, duration up to $3.3 \pm 2.1$~s. 

\begin{figure}
\begin{center}
 \epsscale{0.9}
\plotone{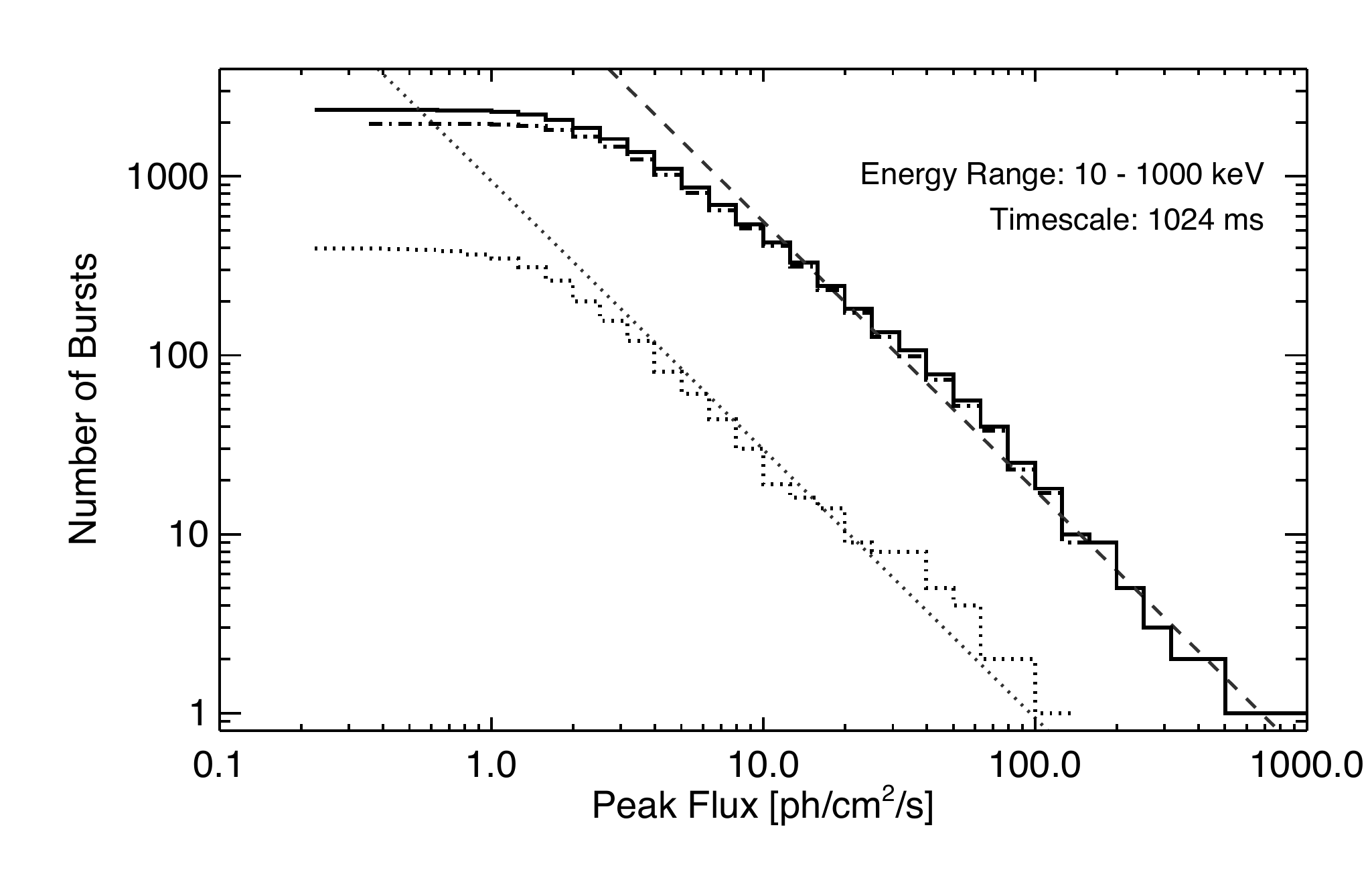}
\plotone{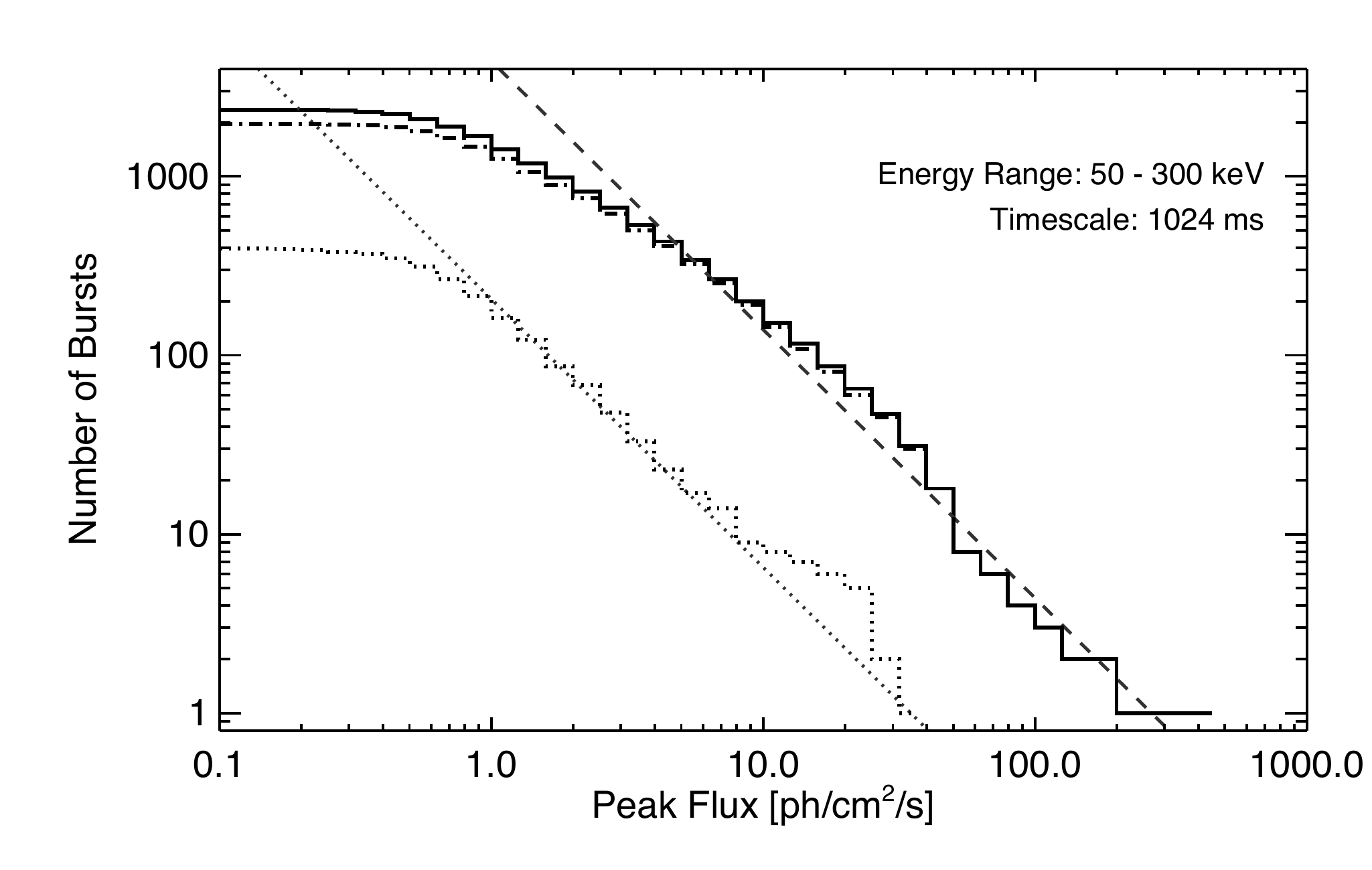}
\caption{\label{fig:pflx_fig} Integral distribution of GRB peak flux on the 1.024~s timescale. Energy ranges are 10--1000~keV (upper plot) and 50--300~keV (lower plot). Distributions are shown for the total sample (solid histogram), short GRBs (dots) and long GRBs (dash-dots), using $T_{90} = 2$~s as the distinguishing criterion. In each plot a power law with a slope of $-3 / 2$ (dashed line) is drawn to guide the eye.}
\end{center}
\end{figure}

\begin{figure}
\begin{center}
 \epsscale{0.9}
\plotone{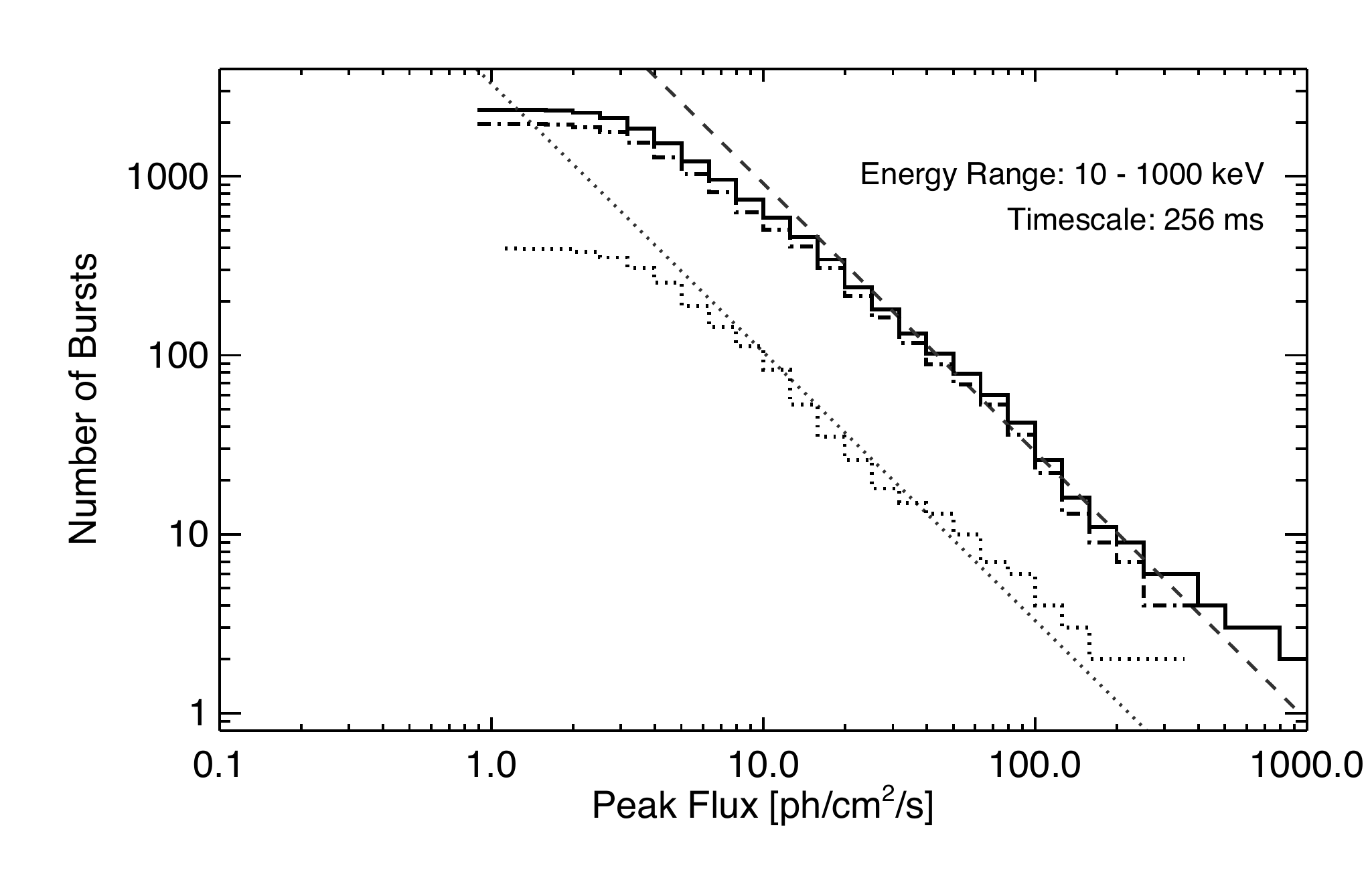}
\plotone{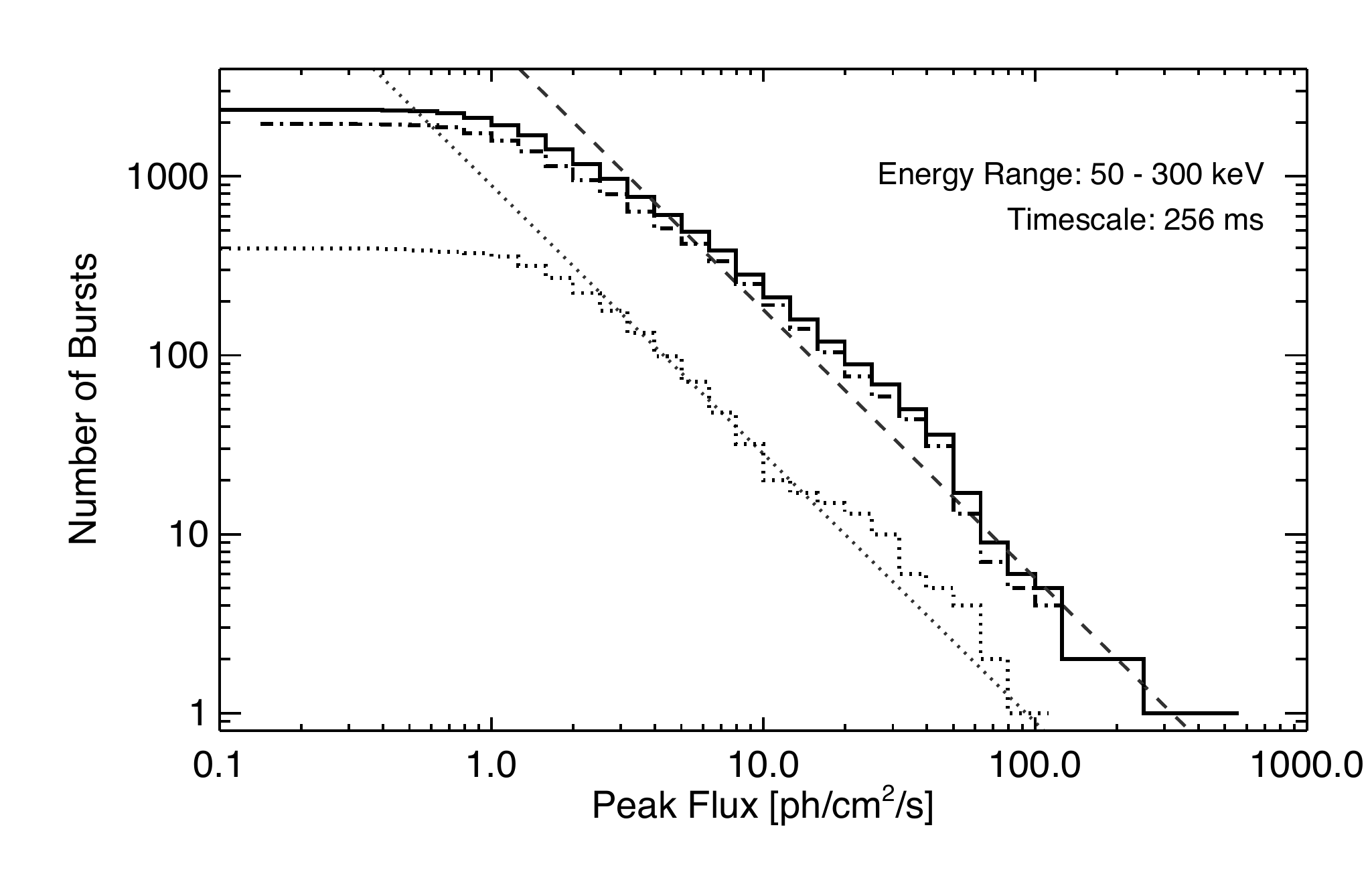}
\caption{\label{fig:pf256_fig} Same as Figure~\ref{fig:pflx_fig}, except on the 0.256~s timescale.}
\end{center}
\end{figure}

\begin{figure}
\begin{center}
 \epsscale{0.9}
\plotone{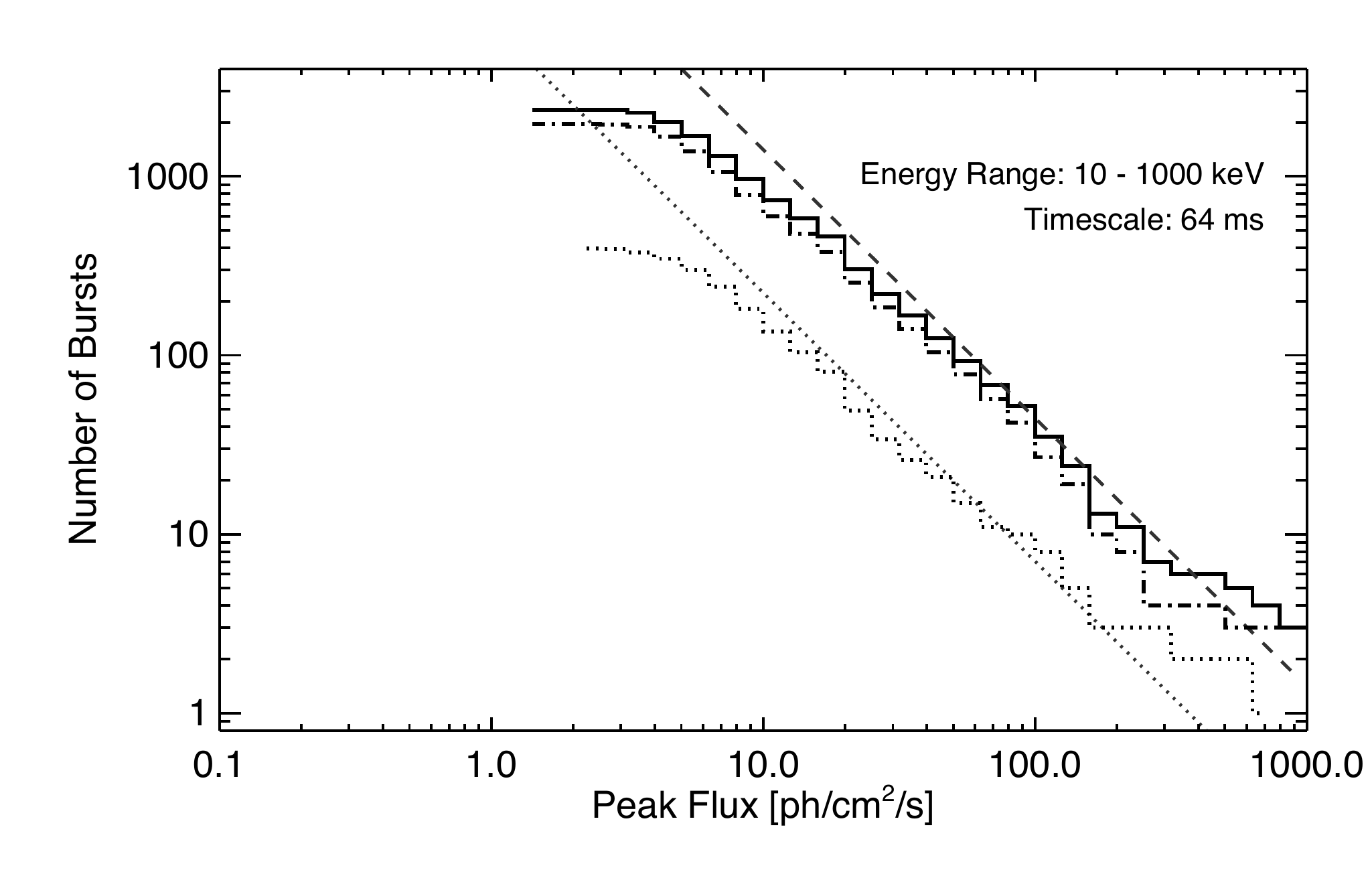}
\plotone{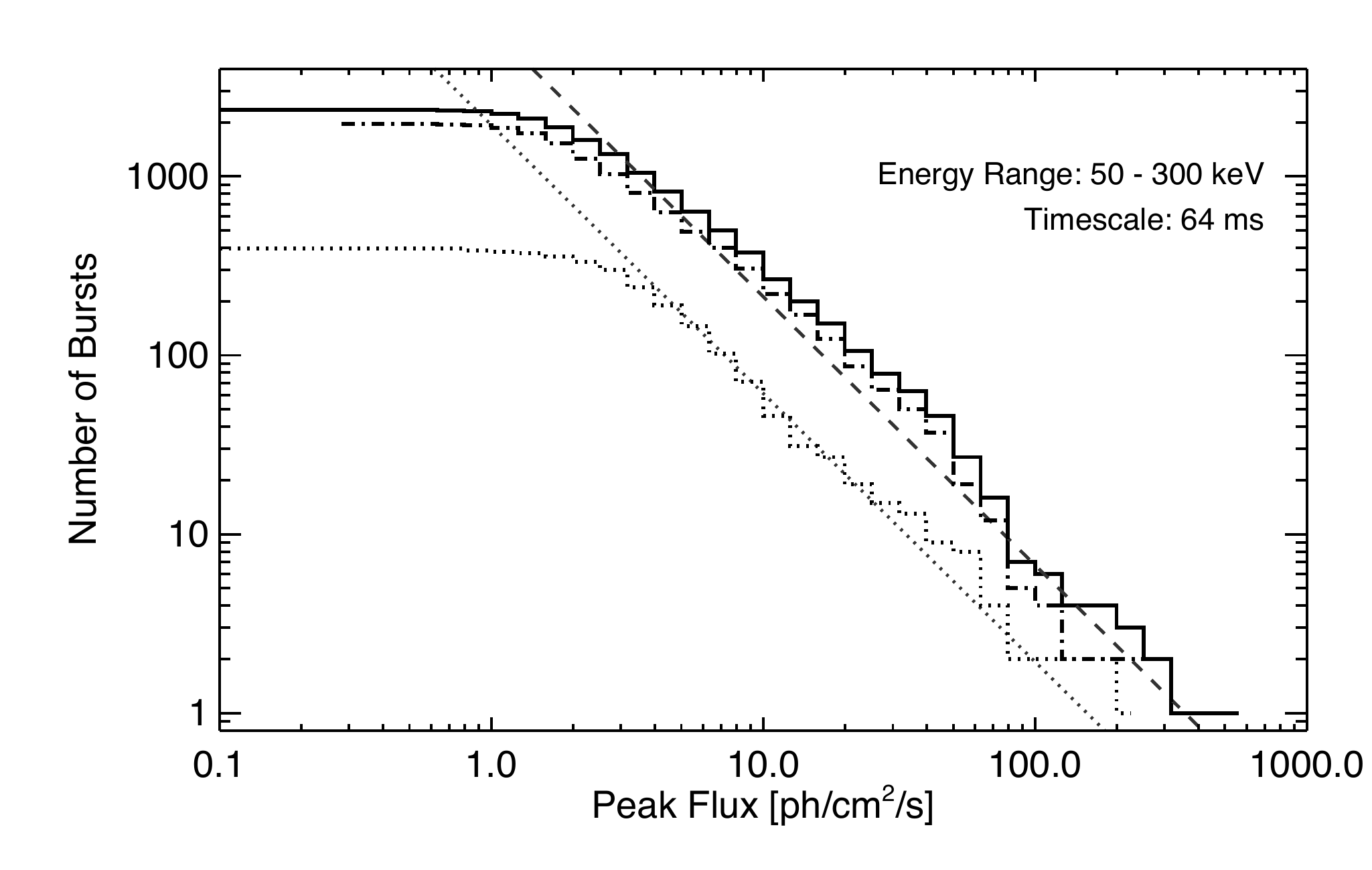}
\caption{\label{fig:pf64_fig}Same as Figure~\ref{fig:pflx_fig}, except on the 0.064~s timescale.}
\end{center}
\end{figure}

\begin{figure}
\begin{center}
 \epsscale{0.8}
\plotone{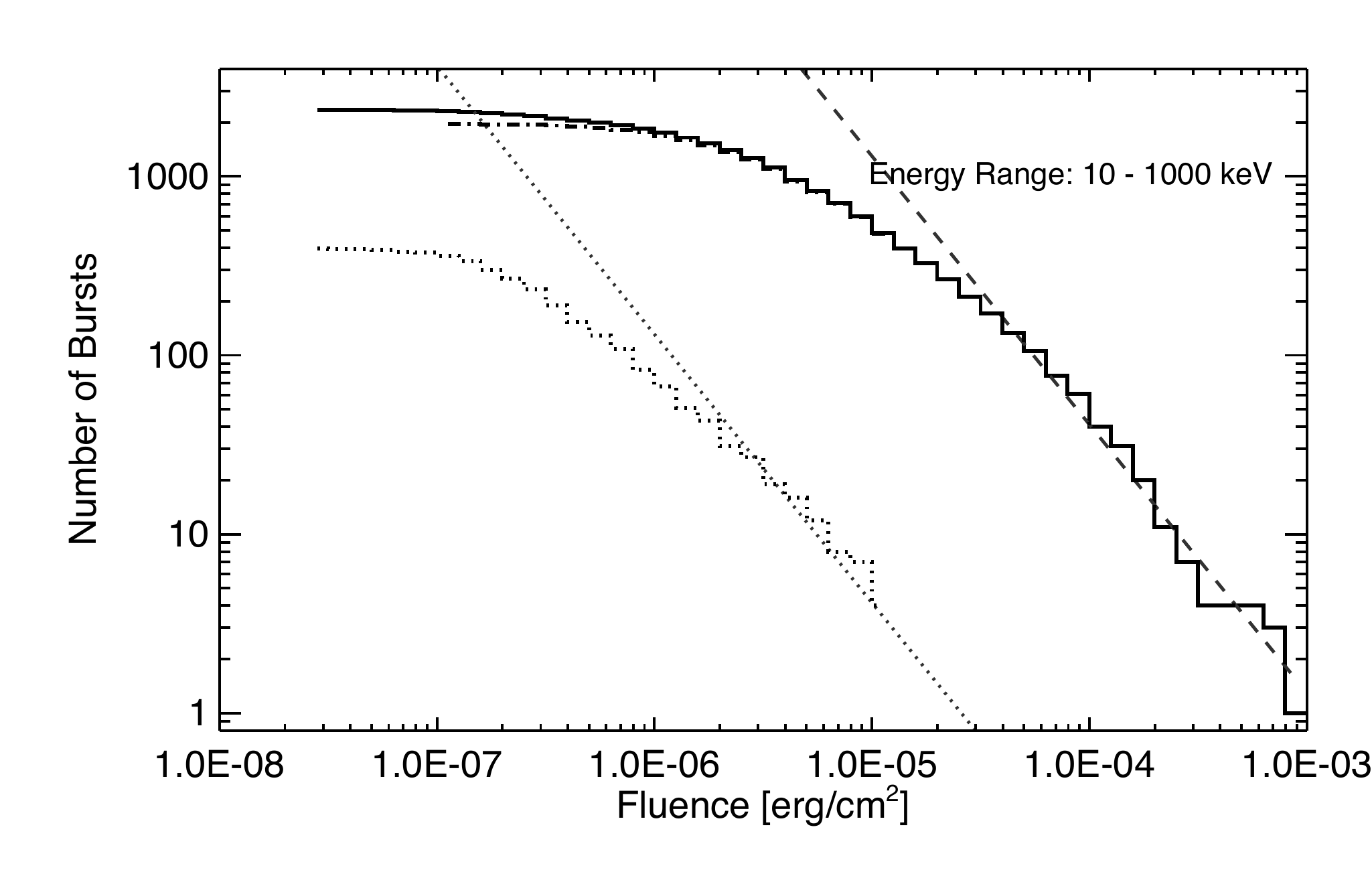}
\plotone{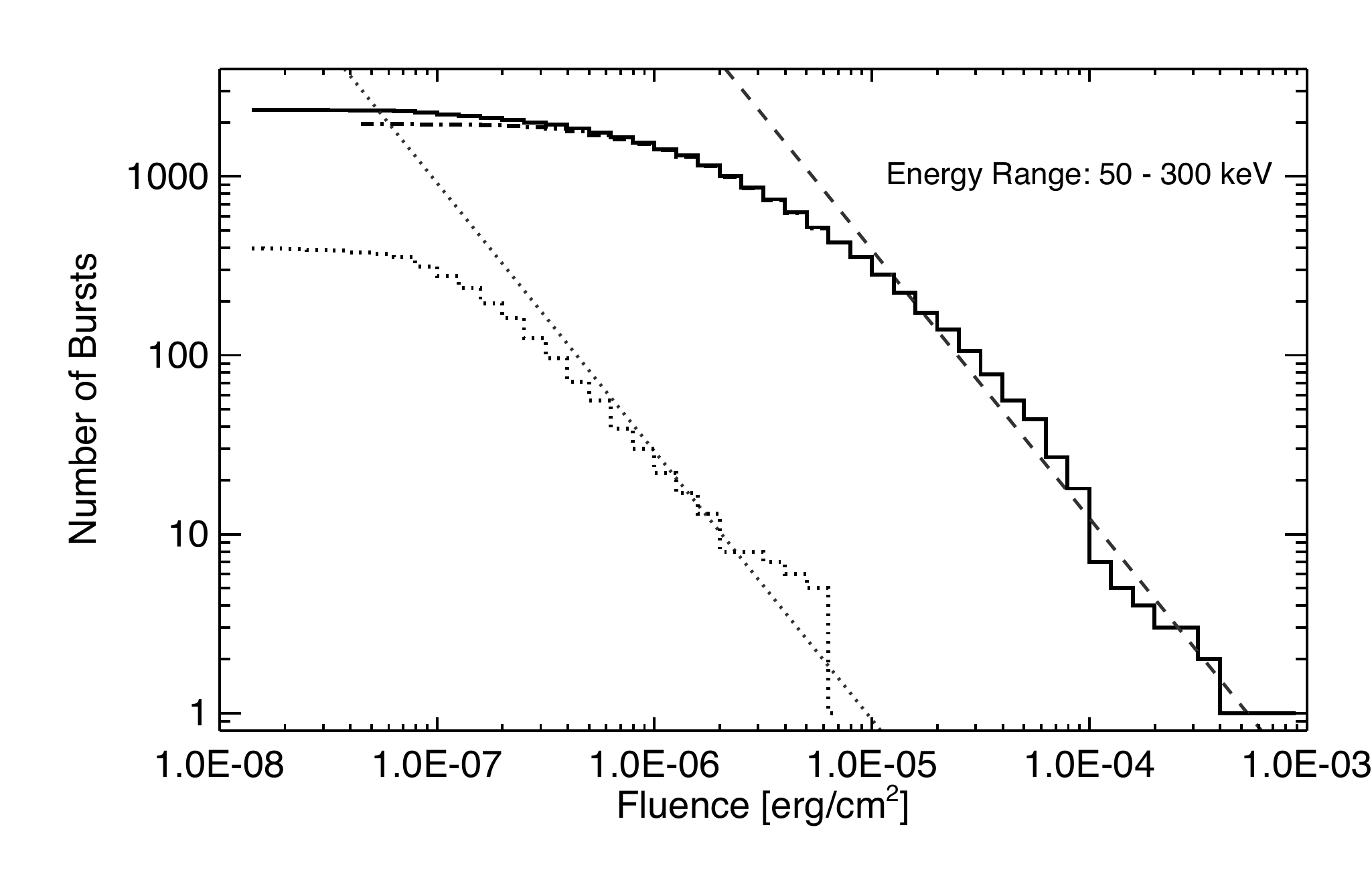}
\caption{\label{fig:flu_fig} Integral distribution of GRB fluence in two energy ranges: 10--1000~keV (upper plot) and 50--300 keV (lower plot). Distributions are shown for the total sample (solid histogram), short GRBs (dots) and long GRBs (dash-dots), using $T_{90} = 2$~s as the distinguishing criterion. In each plot a power law with a slope of $-3 / 2$ (dashed line) is drawn to guide the eye.}
\end{center}
\end{figure}

Integral distributions of the peak fluxes observed for GRBs in the first decade are shown in Figure \ref{fig:pflx_fig} for a 1.024~s timescale and in Figures \ref{fig:pf256_fig} and \ref{fig:pf64_fig} for the shorter  0.256~s and 0.064~s timescales, each separately for short and long GRBs.
The conclusions drawn in previous catalogs regarding the shape of the integral distributions are strengthened. For long GRBs the deviation from the ${-3}/2$ power law, expected for spatially homogeneous GRBs in Euclidean space, occurs well above the GBM threshold at a flux value of $\sim 10$ ph cm$^{-2}$ s$^{-1}$. For short events the GBM data appear consistent with a homogeneous spatial distribution down to peak flux values around 1 ph cm$^{-2}$ s$^{-1}$ (50--300 keV), below which instrument threshold effects become dominant. 
The integral fluence distributions for the two energy intervals are shown in Figure \ref{fig:flu_fig}.

\section{Summary}
\label{sec:Summary}

The fourth \Fermi-GBM Gamma-Ray Burst Catalog comprises a list of 2356 cosmic
GRBs that triggered GBM between 2008 July 12 and 2018 July 11. 
It provides actualized tables and standard analysis results of the full 10 year sample of GBM-triggered GRBs and continues the reporting on exceptional instrument operation conditions; as such it serves as a standard  database and reference for catalog-based follow-up analysis. 

Standard representations of the catalog quantities and analysis results such as the sky distribution of  GBM-triggered GRBs locations, the histograms of GRB \tn- and \tf--duration's and the integral distributions of GRB peak fluxes and fluences now resemble the known characteristics for the now large sample. However, for the presentation of the GRB duration versus hardness and the  hardness distribution itself a more realistic presentation including the parameter uncertainties was introduced. This representation shows a less clear separation of the two most commonly anticipated constituents, the short and long GRB classes. It suggests that about a quarter of the whole GBM GRB sample is due to short GRBs, which is significantly larger compared to the fraction derived when applying the conventional division at 2 s (17 \% short GRBs).  

\acknowledgments
Support for the German contribution to GBM was provided by the Bundesministerium f{\"u}r Bildung und Forschung (BMBF) via the Deutsches Zentrum f{\"u}r Luft und Raumfahrt (DLR) under grant number 50 QV 0301. The USRA coauthors gratefully acknowledge NASA funding through contract NNM13AA43C. The UAH coauthors gratefully acknowledge NASA funding from cooperative agreement NNM11AA01A. E.B. and C.M. are supported by an appointment to the NASA Postdoctoral Program, administered by the Universities Space Research Association under contract with NASA. D.K., C.A.W.H., and C.M.H. gratefully acknowledge NASA funding through the \Fermi-GBM project. 

\bibliography{references}

\clearpage

\clearpage

\clearpage

 \clearpage

\clearpage

\include{main_grb_table_cat4}

\include{duration_table_cat4}

\include{pflux_fluence_table_cat4}

\include{pflux_fluence_b_table_cat4}

\end{document}